\documentclass[12pt, draftclsnofoot,onecolumn]{IEEEtran}
\IEEEoverridecommandlockouts
\usepackage{amsmath}
\usepackage{amsfonts}
\usepackage{bbding}
\usepackage{amssymb}
\usepackage{array}
\usepackage{subfigure}
\usepackage{graphicx}
\usepackage{subfigure}
\usepackage[named]{algo}
\usepackage{psfrag}
\usepackage{xfrac}
\usepackage{booktabs}
\usepackage{stfloats}
\usepackage{cases}
\usepackage[compress]{cite}
\usepackage{hyperref}
\usepackage{bm}
\usepackage{multirow}
\usepackage{amsthm}
\usepackage{stfloats}
\usepackage{setspace}
\usepackage{color}
\usepackage{algorithm}
\usepackage{algorithmicx}
\usepackage{algpseudocode}
\usepackage{amsmath}
\usepackage{amsmath,amssymb,amsfonts,mathrsfs,bm}
\usepackage{cleveref}

\newcommand{\tabincell}[2]{\begin{tabular}{@{}#1@{}}#2\end{tabular}}
\makeatletter
\renewcommand{\citepunct}{,\penalty\@m\hskip.13emplus.1emminus.1em}
\renewcommand{\citedash}{\hbox{--}\penalty\@m}

\makeatother

\newtheorem{rem}{Remark}

\newtheorem{prop}{Proposition}
\newtheorem{cor}{Corollary}

\allowdisplaybreaks

\begin{document}
\vspace{-8mm}
\title{Size Generalization for Resource Allocation with Graph Neural Networks}
\author{
\IEEEauthorblockN{{Jiajun Wu, Chengjian Sun and Chenyang Yang}}
\IEEEauthorblockA{ \\ Beihang University, Beijing, China \\  Email: \{jiajunwu,sunchengjian,cyyang\}@buaa.edu.cn}
}

\maketitle
\vspace{-10mm}
\begin{abstract}
Size generalization is important for learning wireless policies, which are often with dynamic
sizes, say caused by time-varying number of users. Recent works of learning to optimize resource allocation empirically demonstrate that graph neural networks (GNNs) can generalize to different problem scales.
However, GNNs are not guaranteed to generalize across input sizes. In this paper, we strive to analyze the size generalization mechanism of GNNs when learning permutation equivariant (PE) policies. We find that the aggregation
function and activation functions of a GNN play a key role on its size generalization ability. We take the GNN with mean aggregator, called mean-GNN, as an example to demonstrate a size generalization condition, and interpret why several GNNs in the literature of wireless communications can generalize well to problem scales. To illustrate how to design GNNs with size generalizability according to our finding, we consider power and bandwidth allocation, and suggest to select or pre-train activation function in the output layer of mean-GNN for learning the PE policies.
Simulation results show that the proposed GNN can generalize well to the number of users, which validate our analysis for the size generalization condition of GNNs when learning the PE policies.
\end{abstract}
\vspace{-2mm}

\section{Introduction}
Deep learning has been widely applied for resource allocation \cite{LYproc2020,Spawc2017,Eisen2020,LD2020Mag} with diverse motivations, say improving resource usage efficiency by finding real-time policies from NP-hard problems \cite{Spawc2017} or from the problems without models \cite{Alessio2019Model,LD2020Mag}. Intelligent resource allocation has been envisioned as one of the most important features of five-generation beyond and six-generation wireless communications \cite{RMPHY2020,Kato2020}.

A core challenge in designing learning-based wireless resource allocation comes from dynamic propagation environments and fluctuating traffic load. For example, wireless channels and number of users may change quickly. When the number of users vary, the scale of a resource allocation problem and hence the input size of a deep neural network (DNN) for optimizing the problem should be changed accordingly.  While online learning allows a DNN to make decisions over time-varying channels, it is unable to deal with the dynamic caused by time-varying problem scales. Therefore, designing size generalizable DNNs is critical when applying deep learning for resource allocation, which allows the reuse of DNN without the need of retraining or redesigning. If a DNN is unable to generalize to the size of its input, e.g., fully-connected neural network (FNN), the DNN has to be retrained whenever the problem scale changes unless the DNN is trained using the samples with all possible sizes \cite{Mitchell1980}, which incurs non-affordable training complexity for large scale wireless systems.

A promising way to enable generalization is embedding inductive biases into the structure of DNNs. Inductive biases are the constraints imposed on the functions represented by a DNN, which come from the prior knowledge for a class of problems before using data  samples \cite{Mitchell1980}. Mismatched inductive biases may lead to performance degradation due to imposing too strong constraints on the learning process, and appropriate inductive biases can accelerate training without performance loss and find the functions that generalize in a desirable way \cite{GBD1992}. Hence, finding useful inductive biases is important for size generalization.

As a kind of prior knowledge for multivariate functions, permutation equivariance (PE) property has been shown widely existed in wireless resource allocation policies recently \cite{Eisen2020,YS2021JSEC,SCJ2020rank,guo2021learning}. A resource allocation policy can usually be represented as a multivariate function, which is a mapping from environment parameters (say channels of multiple users) to the allocated resources to wireless nodes (say powers to the users). For a policy exhibiting the PE property, the mapping is not affected by the order of nodes. Depending on the considered problem, a resource allocation policy may exhibit one-dimensional (1D)-PE, two-dimensional-PE, joint-PE properties or their combinations \cite{Eisen2020,guo2021learning}. Graph neural networks (GNNs) \cite{keriven2019universal} and permutation equivariant neural networks (PENNs) \cite{Zaheer2017DeepSets} are two kinds of DNNs that are embedded with permutation equivariant inductive biases and hence can learn the PE policies efficiently. In additional to the PE property, GNNs also harness another kind of prior knowledge: the relation among nodes \cite{keriven2019universal}, unless a GNN learns a policy over a complete graph where all nodes are fully connected.


It has been empirically demonstrated in \cite{ML2021TWC,YS2021JSEC,Eisen2020,guo2021learning,ZGIOT} that GNNs can generalize to different graph sizes. In \cite{Eisen2020,YS2021JSEC}, a GNN designed to learn a power control policy in ad hoc or in an interference network was shown able to generalize to $500$ or 1000 transceivers when trained using samples with $50$ transceivers. In \cite{guo2021learning}, a GNN designed to learn a power allocation policy in cellular network was shown able to generalize to the networks with different numbers of users and cells when trained using samples generated with small networks. In \cite{ML2021TWC,ZGIOT}, a GNN designed to learn a link scheduling policy or joint channel and power allocation policy in device-to-device (D2D) communications was shown with good generalization performance to the number of D2D pairs. However, it has been noticed that GNNs are not guaranteed to generalize across the sizes of graphs.
As revealed in \cite{yehudai2021local}, the generalization of GNNs to large graph size is hard for some graph topologies. In fact, the size generalization capability of GNNs is not well-understood. As mentioned in \cite{Eisen2020,YS2021JSEC} without explanation, it is the PE property of GNNs that improves
their size generalizability. As stated in \cite{guo2021learning}, it is the parameter sharing of GNNs that enables good generalization to unseen sizes since the number of trainable parameters of a GNN is independent from the graph size, which is again owing to the PE property. The reason of size generalization ability of GNNs was not mentioned in \cite{ML2021TWC,ZGIOT}.
Though important, it remains unknown when GNNs can generalize to the problem scales, when cannot, and why.

In this paper, we present a small step in understanding size generalization capability of GNNs. To the best of our knowledge, this is the first attempt to reveal the size generalization mechanism in wireless communications. We notice that learning a resource allocation policy with size generalization ability can be regarded as finding the mapping from environment parameters and problem size to the allocated resources with DNN. A straightforward approach to find the mapping is to train a versatile DNN with the samples of all possible sizes, which does not require size generalization. Since training such a DNN is with prohibitive complexity for large scale systems, we resort to the decomposability of a PE policy in the large size regime.
Since the general study for size generalization is extremely difficult, we only analyze case-by-case. To isolate the impact
of the PE property from the inductive bias imposed by relation among nodes, we consider GNNs and policies with 1D-PE property since the policies are learned over a complete graph. In the sequel, we call a 1D-PE policy as PE policy for short. In particular, we take the power and bandwidth allocation policies as example to
demonstrate how to leverage the decomposability for enabling the GNN to generalize to unseen sizes. While GNNs are very flexible, which can use different aggregators and combiners \cite{keriven2019universal}, we consider the commonly used mean, sum, and max  aggregators.
The major contributions are summarized as follows.
\begin{itemize}
	\item We find that aggregation and activation functions impose the inductive bias for size generalization of a GNN. We prove that input-output relation of the GNN with mean aggregator (referred to as mean-GNN for short) and size-independent activation functions (SI-AFs) is invariant to the input size in large size regime, but the GNN with sum or max aggregator does not. We provide a size generalization condition, indicating when a GNN can generalize to the size of a PE policy. From the condition, we can interpret why GNNs used for learning resource allocation policies in existing literatures can generalize well to problem scales.
	\item Inspired by our finding, we suggest a method to improve size generalization, which aligns the inductive bias of a GNN to the size-scaling law of a PE policy by using the mean-GNN with a selected or pre-trained activation function of the output layer. We evaluate the size generalization performance of the proposed method by optimizing the power and bandwidth allocation in ultra-reliable low-latency communications (URLLC) in the scenarios with both large and small number of users.

\end{itemize}

The rest of the paper is organized as follows. In section II, we provide the size generalization condition. In section III, we consider resource allocation policies that are independent from sizes asymptotically, and explain why several GNNs in literature can generalize to the problem scales. In section IV, we consider size-dependent policies, and show how to design the GNNs to enable size generalization. In section V, we evaluate the performance of the proposed GNN for learning the power and bandwidth allocation. In section VI, we provide the conclusion remarks.

\vspace{-0.1mm}
\section{Size Generalization Condition}
In this section, we first introduce the issue of size generalization for learning a resource allocation policy. Then, we show the asymptotic form of the PE policy and the size-scaling laws of the GNNs with mean-, max-, and sum-aggregators. Finally, we show that mean-GNN with SI-AFs can generalize to the size of the PE policy with asymptotic size-invariance property, which implies a condition for size generalization.

\vspace{-2mm}\subsection{Learning a Policy with Different Sizes}
A resource allocation policy for $K$ objects (say users or transceiver pairs) can usually be represented as a multivariate function,
 where the allocated resources (say transmit powers) ${\bf y}$ are the output variables and  the environmental parameters (say channels) ${\bf x}$ are the input variables. Denote ${\bf x} \triangleq [x_{1},...,x_{K}]$, ${\bf y} \triangleq [y_{1},...,y_{K}]$, where $x_k$ and $y_k$ are respectively the input and output variables of the policy for the $k$th object, and $K$ reflects the scale of a resource allocation problem and is referred to as the size of corresponding policy.

In fact, $K$ can also be regarded as an environmental parameter. Then, the resource allocation policy can be represented as ${\bf y}={\bf F}'({\bf x},K)\triangleq [f'_1({\bf x},K),...,f'_K({\bf x},K)]$. If we can learn the function ${\bf F}'({\bf x},K)$ with a DNN, then the DNN can be served as the resource allocation policy with different sizes.
To train such a DNN without the need for size generalization, one can use the samples for ${\bf x}$ with all possible sizes of $K$, which is expensive for large scale systems.

Fortunately, for a PE policy, the multivariate function degenerates into a scalar function of $K$ for each object when $K \to \infty $ as to be shown in the following, i.e., $f'_k({\bf x},K) \rightarrow f(x_k,K )$. This inspires us to learn ${\bf F}'({\bf x},K)$ by learning $f(x_k,K )$ and a size-independent function.

\vspace{-4mm}\subsection{Asymptotic Form of PE Policy}
If the multivariate function ${\bf F}'({\bf x},K)$ is permutation equivariant to ${\bf x}$, i.e., the policy does not change with the order of objects, then it is a PE policy \cite{SCJ2020rank,Eisen2020}. The PE policy satisfies ${\bm \Pi}{\bf y}={\bf F}'({\bm \Pi} {\bf x},K)$, where ${\bm \Pi}$ denotes arbitrary permutation matrix.

In \cite{Zaheer2017DeepSets}, it was proved that permutation invariant functions can be decomposed into continuous outer and inner functions, similar to the Kolmogorov-Arnold representation theorem applicable for arbitrary multivariate continuous function.  By extending the proof in \cite{Zaheer2017DeepSets}, it was proved in  \cite{sannai2019universal}  that the $k$th output variable of any PE policy can be expressed as,\vspace{-1mm}
\begin{align}\label{PE-expre}
y_k=f'_k({\bf x},K) \triangleq \tilde{f} \big(x_k, \sum_{j\neq k} \phi(x_j) \big), \ \ \forall k,
\end{align}\vspace{-1mm}
where $\phi(\cdot)$ and $\tilde{f}(\cdot)$ are continuous functions independent from $k$.

The following proposition shows the asymptotic form of \eqref{PE-expre} for independent and identically distributed (i.i.d.) input variables.

\vspace{-3mm}\begin{prop}\label{DecouplePolicy}
For the PE policy, if $K \to \infty$ and $x_k, k=1,\cdots, K$ are i.i.d., then\vspace{-1mm}
\begin{align}
y_k  =  f(x_k,K ), \label{Define_fk}
\end{align}
where the expression of $f(x_k,K )$ is shown in the appendix.\vspace{-1mm}
\end{prop}\vspace{-3mm}
\begin{proof}
See Appendix \ref{proof P1}
\end{proof}\vspace{2mm}
It indicates that a PE policy asymptotically degenerates into a function of $K$, which maps the input variable to the output variable of each object.

\vspace{-3mm}
\subsection{Impact of Aggregator on the Scaling Law of a GNN with Size}
We first consider a GNN with mean aggregator, i.e., \emph{mean-GNN}, for learning the PE policy with $K$ objects. Denote the output of the GNN as $\hat{\bf y} \triangleq [\hat{y}_{1},...,\hat{y}_{K}]$, and the output of the $l$th layer as $[{\bf h}_{1}^l,...,{\bf h}_{K}^l]^T$. The relation between ${\bf h}_{k}^l$ and ${\bf h}_k^{l-1}$ of the mean-GNN can be expressed as
\begin{align}
{\bf h}_{k}^l\!=\!\sigma^l({\bf U}^{l}{\bf h}_{k}^{l-1}\!+\!{\bf V}^{l}(\frac{1}{K}\sum_{j \neq k} {\bf h}_{j}^{l-1})\! +\! {\bf c}^l), \  l\!=\!1,...,L+1, \label{Out-hidd}
\end{align}
where $\sigma^l(\cdot)$ is the activation function in the $l$th layer, ${\bf U}^{l}$, ${\bf V}^{l}$ and ${\bf c}^l$ are the trainable parameters of the $l$th layer, and $L$ is the number of hidden layers. ${\bf h}_{k}^{0} = {x}_{k}$, and ${\bf h}_{k}^{L+1} =\hat{y}_{k}$.

The term $\frac{1}{K}\sum_{j \neq k} {\bf h}_{j}^{l-1}$ in \eqref{Out-hidd} is an average over ${\bf h}_{j}^{l-1}, j \neq k$. The average is a pooling operation, which is called mean aggregation in literature.
In order to impose PE inductive biases on GNNs to learn PE policies efficiently, the aggregation function should satisfy commutative law, hence it can also be other operations such as summation and maximization.

The following proposition shows the scaling law of mean-GNN for  i.i.d. input features. It indicates that the $k$th output of the mean-GNN well-trained for $K$ objects depends on ${x}_{k}$ and $K$ but is independent from ${x}_{k'}, k'\neq k$ when $K \rightarrow \infty$.

\vspace{-2mm}\begin{prop}\label{GNN-prop}
For the mean-GNN, if $K \to \infty$  and $x_k, k=1\cdots, K$ are i.i.d., then
\begin{align}
\hat{y}_{k} = \hat q(x_k,K), \label{OutGNN}
\end{align}
where the expression of $\hat q(x_k,K)$ is shown in the appendix.
\end{prop}\vspace{-2mm}
\begin{proof}
See Appendix \ref{iid-hidd}.
\end{proof}

Denote the probability distribution function of the features in ${\bf x}$ of size $K$ as $p^K(x)$.
The following proposition shows an asymptotic size-invariance property of the mean-GNN.
\begin{prop}\label{GNN-prop}
For the mean-GNN, if $ p^K({x})=p^{K'}({x})$ for any $K \neq K'$, when $K, K' \to \infty$ and $\sigma^l(\cdot), l=1,\cdots, L+1$ do not depend on $K$, then the relation between its input and output of each object is invariant to the input size, i.e.,
\vspace{-2mm}\begin{align} \label{Prop-GNN}
\hat y_k =\hat q(x_k,K) = \hat q(x_k,K') \triangleq \hat q(x_k).
\end{align}
\end{prop}\vspace{-2mm}
\begin{proof}
See Appendix \ref{App-prop1}.
\end{proof}

The proposition means that if the input features are with identical distribution for different input sizes meanwhile the activation function in each layer is independent from $K$, then the functions able to be represented by the mean-GNN do not depend on $K$. Most commonly used activation functions do not depend on the input size, e.g., \texttt{Relu} and its variants, \texttt{sigmoid}, \texttt{Softplus} and \texttt{Tanh}. One exception is \texttt{Softmax}.

However, the GNN with max-aggregator or sum-aggregator is not size invariant.
\vspace{-2mm}\begin{prop}\label{GNN-prop3}
For the GNN with maximization or summation as aggregation function, if $K \to \infty$  and $x_k, k=1\cdots, K$ are i.i.d., then $\hat{y}_{k} = \hat q(x_k,K)$.
However, if $ p^K({x})=p^{K'}({x})$ for any $K \neq K'$, $K,K' \rightarrow \infty$ and $\sigma^l(\cdot), l=1,\cdots, L+1$ do not depend on $K$, its input-output relation is no longer invariant to the input size.
\end{prop}\vspace{-2mm}
\begin{proof}
See Appendix \ref{Pool-max-sum}.
\end{proof}

\vspace{-3mm}\subsection{Size Generalization Condition}
When the mean-GNN is well-trained for a PE policy with $K$ objects, then $\hat q(x_k,K) \approx  f(x_k,K)$ if $K \rightarrow \infty$ and features are i.i.d. according to Propositions 1 and 2 and the universal approximation theorem for PE functions  \cite{sannai2019universal}. Further considering Proposition 3, we can obtain the following corollary for the PE policy with size-invariance property.
\vspace{-2mm}\begin{cor}\label{Dim-cond}
If a PE policy is asymptotically invariant to the size $K$, i.e.,
\begin{align}\label{PE-cond}
y_k = f(x_k,K) = f(x_k,K') \triangleq f(x_k), ~K \neq K', K,K' \rightarrow \infty,
\end{align}
then the mean-GNN  with SI-AF in each layer trained for learning the PE policy with $K$ objects can generalize to the PE policy with $K'$ objects when $K,K' \rightarrow \infty$.
\end{cor}

\begin{figure}[!htbp]
    \centering
    \includegraphics[width=12cm,height=6cm]{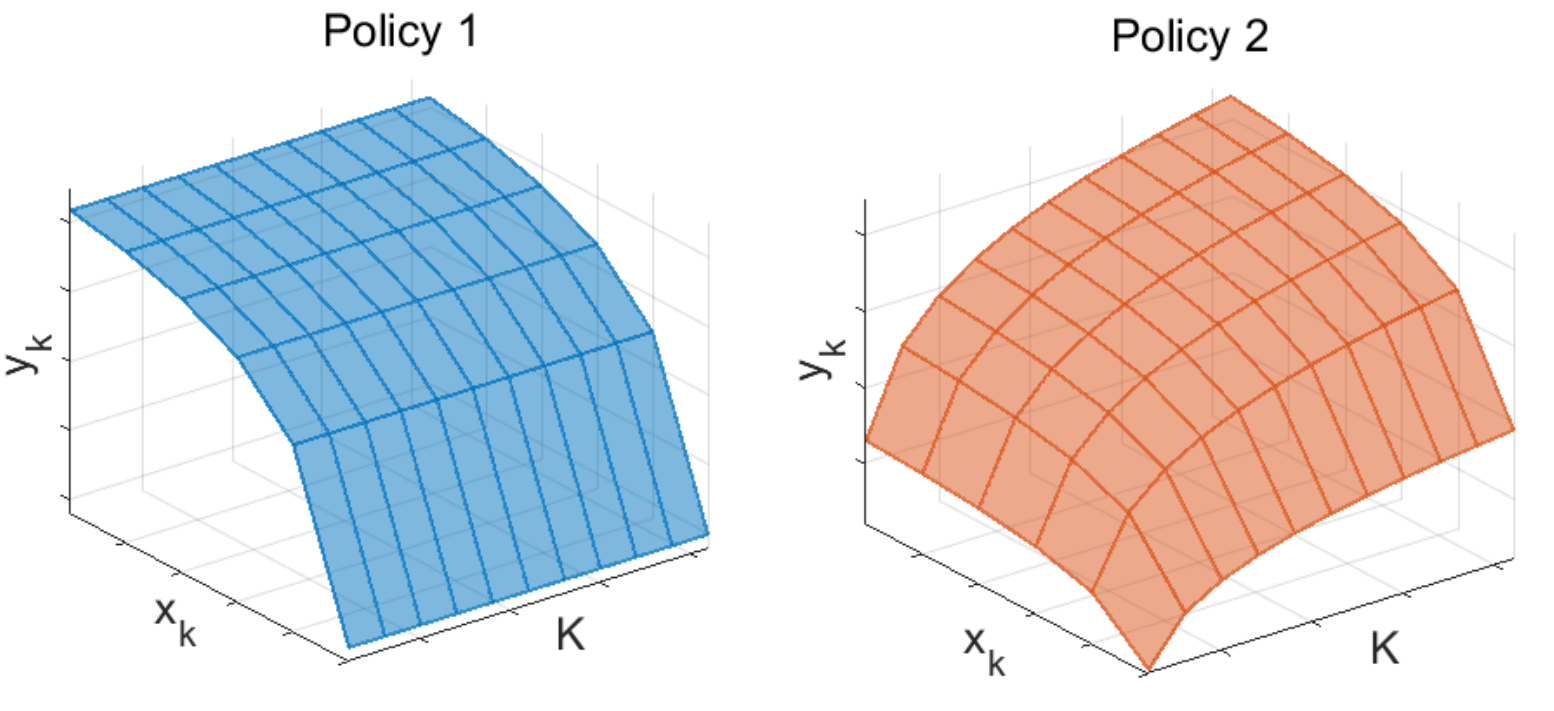}\vspace{-3mm}
    \caption{ Illustration of $y_k = f(x_k,K)$ of two policies versus $K$.}
    \label{Polciy-ill}\vspace{-5mm}
\end{figure}

To help understand the PE polices with the size-invariance property, we show $y_k = f(x_k,K)$ of two toy example policies in Fig. \ref{Polciy-ill}. For any given input variable $x_k$, the output variable $y_k$ of ``Policy 1" does not depend on $K$, while $y_k$ of ``Policy 2" increases with $K$. Thus, ``Policy 1" satisfies the size-invariance property but ``Policy 2" does not.

Asymptotic size-invariance is the simplest size-scaling law. Not all PE policies exhibit the size-invariance property. For a PE policy satisfying \eqref{PE-cond}, the mean-GNN with SI-AFs can generalize to the policy with different sizes, but the GNN with max- or sum-aggregator can not.

This implies a condition for a GNN to learn a PE policy with size generalization ability.
 \vspace{-0.2mm}

\begin{rem}
Size generalization condition: a GNN can generalize to the size of a PE policy if the size-scaling law of the GNN aligns with the size-scaling law of the PE policy.
\end{rem}\vspace{-3mm}

\vspace{-2mm}\section{Size Generalization for Policies with Size-invariance Property}
In this section, we provide three size-invariant PE policies, from which we strive to interpret why existing GNNs used to learn these policies can generalize to the number of objects. We start from a simple policy with closed-form expression to help understand the size-invariance property. We then consider two  resource allocation policies without closed-form expressions, to explain why the GNNs in \cite{YS2021JSEC,ML2021TWC} have size generalization ability for learning these policies.

\vspace{-3mm}\subsection{Power Allocation Policy with Closed-form}
Consider a downlink frequency division multiple access system, where a single-antenna base station (BS) serves $K$ single-antenna users over bandwidth $B$. The power allocation is optimized to minimize the total transmit power under the quality-of-service (QoS) constraint as follows,
\begin{align}
\min_{P_k} \ \ & \sum_{k=1}^K P_k  \label{P5}\\
s.t. \  \ & B\log(1+\frac{ P_k g_k }{N_0B}) \geq s_0,  \tag{\ref{P5}a} \label{P5_a} \\ & P_k \geq 0, \tag{\ref{P5}b} \label{P5_b}
\end{align}
where  $P_k$ is the power  allocated to the $k$th user, $g_k$ is the channel gain of the user, $N_0$ is the single-side noise spectral density, and $s_0$ is the minimal data rate required by each user.


The problem is convex and its global optimal solution can be obtained from the Karush-Kuhn-Tucker (KKT) conditions as $P_k^* =  \frac{N_0}{g_k}B(2^{ \frac{s_0 }{ B}  }-1)$.

The power allocation policy obtained from \eqref{P5} can be expressed as ${\bf p}^*={\bf F}_1({\bf g})$, where ${\bf p}^*=[P^*_{1},...,P^*_{K}]$ and ${\bf g}=[g_{1},...,g_{K}]$. Since the multivariate function ${\bf F}_1({\bf g})$ is not affected by the order of users, the policy is a PE Policy.

${\bf F}_1({\bf g})$ is independent of $K$, since the problem can be decoupled into multiple problems each for a single-user. In other words, the power allocation policy is size-invariant for all values of $K$, and hence can be learned by the mean-GNN with SI-AFs with size generalization ability.

\vspace{-2mm}\subsection{Power Control or Link Scheduling Policy in Interference Networks}
Consider a power control policy for an interference network consisting of $K$ single-antenna transceiver pairs in \cite{Spawc2017}. The transmit power is optimized to maximize the sum rate under the power constraint at each transmitter as follows,
\begin{align}
\label{P2-case1}
\max_{P_k} \ \ & \sum_{k=1}^K  \log \left( 1+\frac{ P_{k} {g}_{k,k}  }{\sum_{j\neq k} P_{j} {g}_{k,j}+\sigma_0 } \right) \\ s.t. \ \ & 0 \leq P_{k} \leq P_{\rm max}, \nonumber
\end{align}
where $P_k$ is the transmit power at the $k$th transmitter, ${g}_{k,j}$ is the channel gain from the $j$th transmitter to the $k$th receiver, $\sigma_0$ is the noise power, and $P_{\rm max}$ is the maximal transmit power of each transmitter.

The optimal power control policy can be expressed as ${\bf p}^*={\bf F}'_2({\bf G},K)$, where ${\bf p}^*=[P^*_{1},...,P^*_{K}]$, and ${\bm G}=[{g}_{k,j}]_{k,j=1}^K$ is the channel matrix whose element in the $k$th row and $j$th column is ${g}_{k,j}$. Since ${\bf F}'_2({\bf G},K)$ is not affected by the order of the transceiver pairs, the policy is a PE Policy.

The problem in \eqref{P2-case1} is without closed-form solution. Numerical solution can be obtained via a weighted sum mean-square error minimization (WMMSE) algorithm, where the solution is binary (i.e., $P_k \in \{0,P_{\rm \max}\}$)  in most cases \cite{Spawc2017}.\footnote{Under Rayleigh fading channel and $P_{\rm max}=1, \sigma_0=1$ as considered in \cite{Spawc2017}, the ratio of the numerical solution satisfying $P_k\leq 10^{-3}$ or $P_k\geq 1-10^{-3}$ is larger than $95\%$.}
In what follows, we numerically show the asymptotic size-invariance property of this complex policy obtained from the WMMSE algorithm.

\begin{figure}[htbp!]
    \centering
    \subfigure[$\Pr\{P_k^*=P_{\rm \max}\}$ of the complex policy with size-invariance property.]{
    \begin{minipage}{5.2cm}
    \centering
    \includegraphics[height=3.9cm,width=5.2cm]{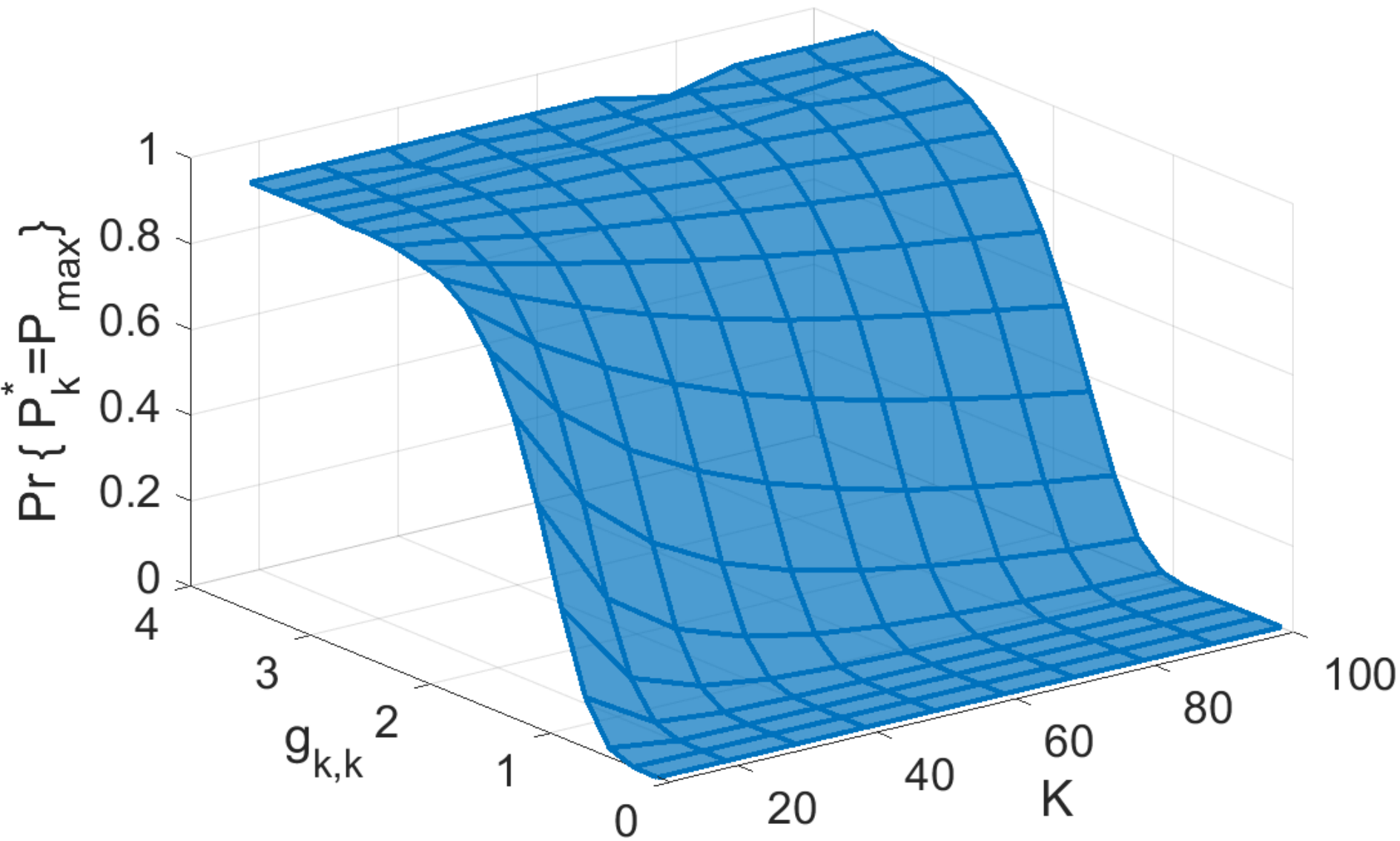}
    \end{minipage}}
    \subfigure[Average $\Pr\{P_k^*=P_{\rm \max}\}$ of the complex policy with size-invariance property. ]{
    \begin{minipage}{5.2cm}
    \centering
    \includegraphics[height=3.9cm,width=5.2cm]{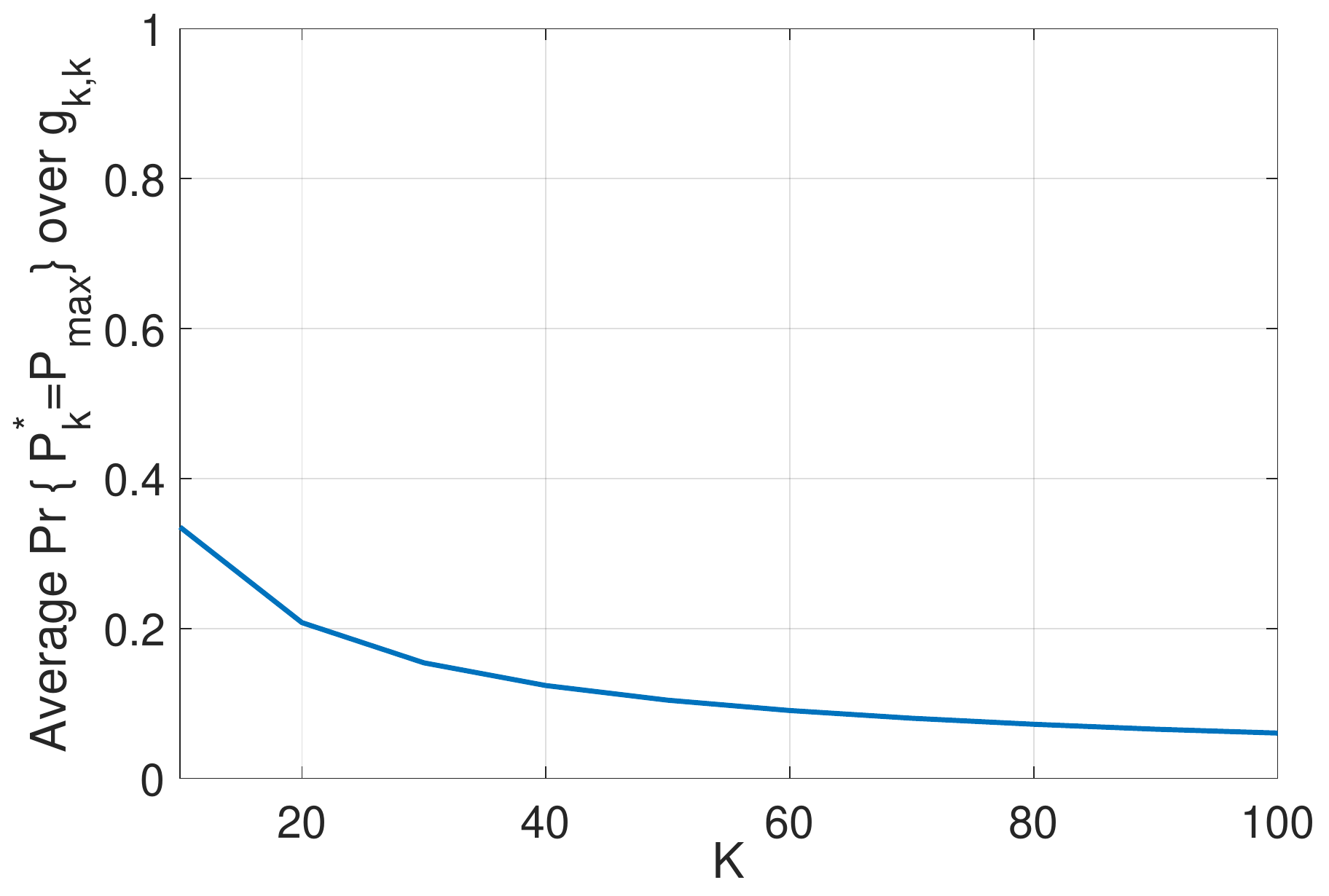}
    \end{minipage}}
    \subfigure[Statistics of $P_k^*$ of the simple policy without size-invariance property. ]{
    \begin{minipage}{5.2cm}
    \centering
    \includegraphics[height=3.9cm,width=5.2cm]{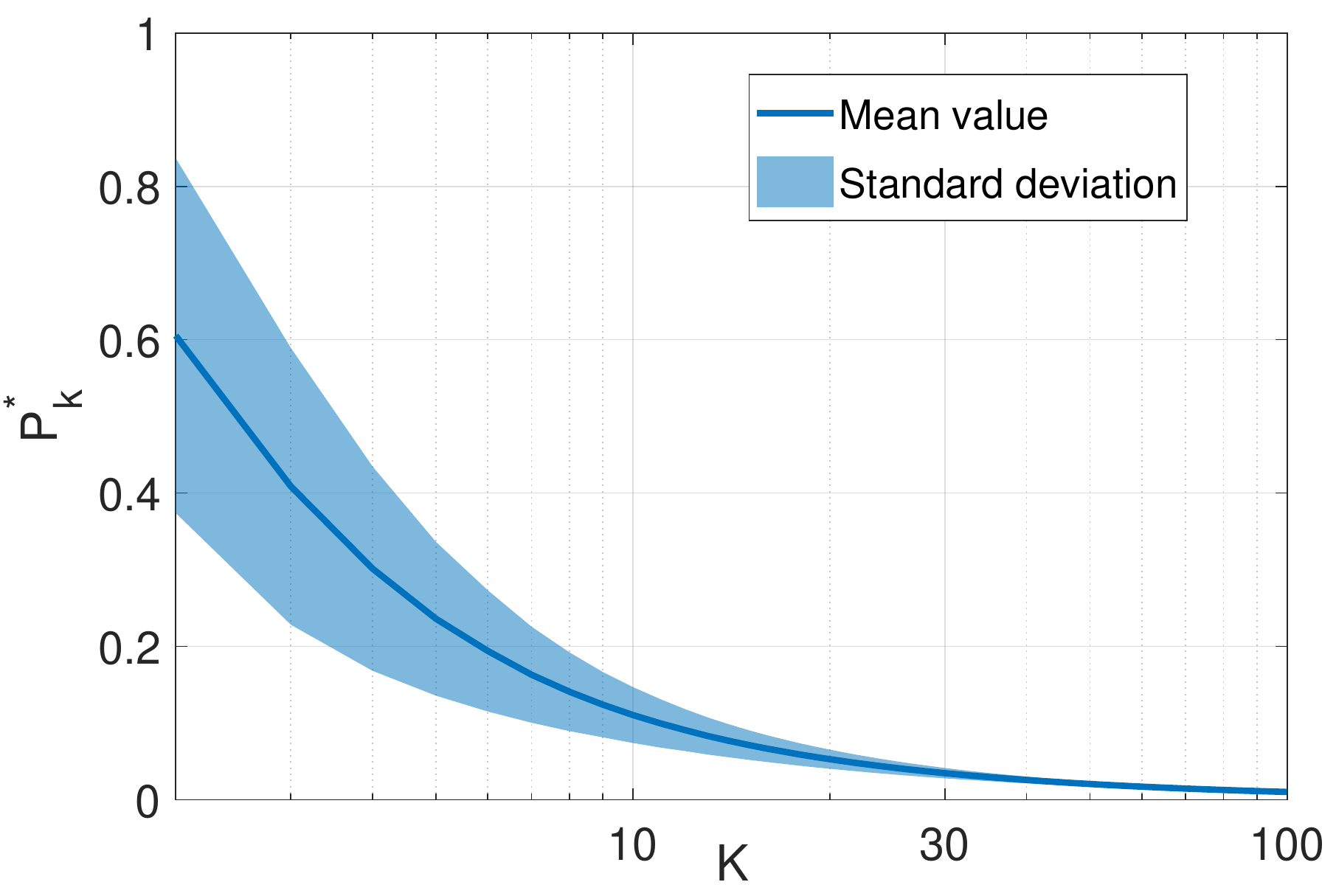}
    \end{minipage}}
    \vspace{-2mm}
    \caption{Behavior of PE policies  with or without the asymptotic size-invariance property versus $K$. }
    \label{Polciy-IC}\vspace{-5mm}
\end{figure}

Considering that $P_k^*$ also depends on the channel gains of interference links (i.e., $g_{k,j}$ and $g_{j,k}$, $j\neq k$) when $K$ is small, we show the probability of $P_k^*=P_{\rm \max}$, i.e., $\Pr\{P_k^*=P_{\rm \max}\}$, when $g_{k,k}$ is given for each value of $K$. We set $K=[10,20,...,100]$ and generate 100,000 realizations of ${\bm G}$ for each value of $K$. Since it is hard to find the same value of $g_{k,k}$ in different realizations, $\Pr\{P_k^*=P_{\rm \max}\}$ is obtained by counting the frequency when $g_{k,k} \in (g_{k,k}-\delta,g_{k,k}+\delta)$, where $\delta$ is a small value and is set as 0.1. We consider Rayleigh fading channel. The result is shown in Fig. \ref{Polciy-IC}(a).
We can see that $\Pr\{P_k^*=P_{\rm \max}\} \approx 1$ for $g_{k,k}\geq 2$ and  $\Pr\{P_k^*=P_{\rm \max}\} \approx 0$ for $g_{k,k}\leq 1$. Moreover, $\Pr\{P_k^*=P_{\rm \max}\}$ is almost invariant to $K$ when $K \geq 50$. To see this more clearly, we provide the mean value of $\Pr\{P_k^*=P_{\rm \max}\}$ by taking the average over $g_{k,k}$  in Fig. \ref{Polciy-IC}(b), which changes slowly with $K$ when $K$ is large.
This indicates that the policy obtained from problem \eqref{P2-case1} is asymptotical  size-invariant with high probability.
\vspace{-0.2mm}

In \cite{ML2021TWC}, link scheduling for D2D systems is optimized to maximize the sum rate as follows,
\begin{align}
\label{P2-D2D-case1}
\max_{\rho_k} \ \ & \sum_{k=1}^K   \log \left( 1+\frac{ \rho_k P_{k} {g}_{k,k}  }{\sum_{j\neq k} \rho_j P_{j} {g}_{k,j} +\sigma_0 } \right) \\ s.t. \ \ &  \rho_k \in \{0,1 \}, \nonumber
\end{align}
where $\rho_k=1$ if the $k$th D2D pair is scheduled and $\rho_k=0$ otherwise.
Since the objective function has the same form as that in \eqref{P2-case1} and the constraint is a special case of that in \eqref{P2-case1}, the optimal policy obtained from \eqref{P2-D2D-case1} is also asymptotical  size-invariant with high probability.

In \cite{ML2021TWC}, the optimal policy obtained from problem \eqref{P2-D2D-case1} is learned by graph embedding (can be regarded as a GNN), where the aggregation function is summation. In \cite{YS2021JSEC}, the optimal policy obtained from problem \eqref{P2-case1} is learned by a GNN, where the aggregation function is maximization.
According to Proposition \ref{GNN-prop3}, the input-output relation of the GNN with sum- or max-aggregator is not invariant to the input size.
Nonetheless, the GNNs have been demonstrated with good size generalization performance in \cite{YS2021JSEC,ML2021TWC}. This is because
each output of the GNN in \cite{ML2021TWC} is followed by FNN for classification, and the output layer of the GNN in \cite{YS2021JSEC} behaves like a classifier, in order for learning a binary solution. Although the hidden output of the GNN in \cite{YS2021JSEC} or \cite{ML2021TWC} is not invariant to the input size, the operation in the output layer  for classification makes $I(\hat{q}(x_k,K)>\eta) = I(\hat{q}(x_k,K')>\eta)$ holding with high probability, where $I(\cdot)$ denotes the indicator function and $\eta$ is the threshold for classification. This suggests that when the size-scaling law of a GNN does not match with the size-scaling law of a PE policy, further processing the output may enable the GNN generalizable to the PE policy with different sizes.

\vspace{-2mm}
\section{Size Generalization for Policies without Size-invariance Property} \label{Case2}
In this section, we consider PE policies without size-invariance property, and demonstrate how to satisfy the size generalization condition in Remark 1 by designing mean-GNNs for learning these policies. We first consider power allocation and bandwidth optimization policies where the power allocation policy is with closed-form expression. Despite that it is unnecessary to learn the policy in practice, the goal to provide this policy is for illustrating how to satisfy the size generalization condition by selecting an activation function for the output layer of mean-GNN with SI-AFs in hidden layers. Then, we consider power and bandwidth allocation in URLLC, whose solution is unable to be found via numerical algorithms. We proceed to propose a method to learn the policy with mean-GNN, which can generalize to different values of $K$.

\vspace{-3mm}\subsection{Power Allocation and Bandwidth Optimization Policy}\label{simply-not}
We still consider the downlink frequency division multiple access system, where a single-antenna BS serves $K$ single-antenna users with maximal transmit power $P_{\rm max}$. Now we optimize the bandwidth (equally allocated to each user) together with power allocation to minimize the total bandwidth under both the QoS constraint and the power constraint,
\begin{align}
\min_{P_k,B} \ \ & K B  \label{P6}\\
s.t. \  \ &  s_k \triangleq B\log(1+\frac{ P_k g_k }{N_0B}) \geq s_0,   \tag{\ref{P6}a} \label{P6_a} \\ & \sum_{k=1}^K P_k \leq P_{\rm max},  \tag{\ref{P6}b} \label{P6_b} \\ &   P_k \geq 0,B \geq 0.  \tag{\ref{P6}c} \label{P6_c}
\end{align}
It is not hard to prove that this problem is convex.
From the KKT conditions,
we find that the global optimal solution, denoted as $P_k^*$ and $B^*$, satisfies the equality of \eqref{P6_b}, and
\begin{align}
&P_k^* = \frac{1}{g_k} F(B^*), \label{P-case2-1a}\\
&F(B^*) \triangleq N_0B^*(2^{ \frac{s_0 }{ B^*}  }-1). \label{P-case2-1b}
\end{align}

By substituting \eqref{P-case2-1a} into $\sum_{k=1}^K P_k^* = P_{\rm max}$, we have $F(B^*)= \frac{P_{\rm max}}{\sum_{k=1}^K \frac{1}{g_k} }$. By substituting this expression of $F(B^*)$ into \eqref{P-case2-1a}, we can obtain the optimal power allocated to the $k$th user and re-write it in the form of \texttt{softmax} function as
\begin{align}
P_k^* &=P_{\rm max}\frac{\frac{1}{g_k}}{\sum_{k=1}^K \frac{1}{g_k} } =P_{\rm max} \frac{ e^{\ln \frac{1}{g_k}}}{\sum_{k=1}^K e^{\ln \frac{1}{g_k}}}.
\label{P-case2-2}
\end{align}

Denote $F^{-1}(\cdot)$ as the inverse function of $F(B)$ in \eqref{P-case2-1b}. The optimal bandwidth allocated to the $k$th user can be re-written as
\begin{align}
 B^*= F^{-1}(\frac{P_{\rm max}}{\sum_{k=1}^K \frac{1}{g_k} }). \label{BK-case2-2}
\end{align}

Denote the power allocation and bandwidth optimization policies obtained from \eqref{P6} as ${\bf p}^*={\bf F}'_3({\bf g},K)$ and $B^*={ f}'_4({\bf g},K)$, respectively, which are PE policies since the functions ${\bf F}'_3({\bf g},K)$ and ${f}'_4({\bf g},K)$ remain unchanged when the order of the users changes.

From the first equality of \eqref{P-case2-2}, the power allocation policy can be expressed as a composite function of $f_{e1}(g_k)\triangleq \frac{1}{g_k}$ and a function for normalization (i.e., $\frac{{z_k}}{\sum_{k=1}^K {z_k}}$). From the second equality, the policy can also be expressed as a composite function of $f_{e2}(g_k)\triangleq \ln \frac{1}{g_k}$ and the \texttt{softmax} function as
\begin{align}
{\bf F}'_3({\bf g},K)  = &  P_{\rm max} \cdot  \Big[ {\sigma}_{s} \big(f_{e2}(g_1),\{ f_{e2}(g_j)\}_{j \neq 1}\big), \cdots,  {\sigma}_{s} \big(f_{e2}(g_K),\{ f_{e2}(g_j)\}_{j \neq K}\big) \Big]
\nonumber \\ \triangleq &  P_{\rm max} \cdot {\bm \sigma}_s \left(f_{e2}(g_1),\cdots,f_{e2}(g_K)\right),
\label{P-case2-equi}
\end{align}
where $\sigma_s ( z_k, \{z_j\}_{j \neq k} )\triangleq \frac{e^{z_k}}{\sum_{k=1}^K e^{z_k}}$. For this PE policy, the inner function $f_{e1}(g_k)$ or $f_{e2}(g_k)$ is size-invariant, and the outer function  $\frac{{z_k}}{\sum_{k=1}^K {z_k}}$ or $\sigma_s (\cdot)$  is size-dependent.

\vspace{2mm}\subsubsection{Asymptotic Policies}
To show that both policies are not size-invariant, we provide their asymptotic form.

When $K \to \infty$, we have $\frac{P_{\rm max}}{(\sum_{k=1}^K \frac{1}{g_k})} = \frac{1}{K}   \frac{P_{\rm max}}{(\sum_{k=1}^K \frac{1}{g_k})/K} $ $\to \frac{1}{K}   \frac{P_{\rm max}}{ \mathbb{E}(\frac{1}{g_k}) } $. Then, the optimal solution in large $K$ regime can be expressed as

\begin{align}\label{P-case2-3}
P_k^* \to \frac{1}{K}   \frac{P_{\rm max} }{ g_k\mathbb{E}(\frac{1}{g_k})} \triangleq f_3(g_k, K), ~~
B^* \to F^{-1} ( \frac{1}{K}   \frac{P_{\rm max}}{ \mathbb{E}(\frac{1}{g_k}) } )  \triangleq f_4(g_k, K).
\end{align}

In order to show how large the value of $K$ is sufficient for $P_k^*$ to be solely dependent on $g_k$ and $K$ as shown in \eqref{P-case2-3}, we provide the values of $P_k^*$ versus $K$ for a given value of $g_k$  in Fig. \ref{Polciy-IC}(c). Since $P_k^*$ also depends on the channel gains of other users (i.e., $g_j, j \neq k$) when $K$ is small as shown in \eqref{P-case2-2}, we show the mean value and standard deviation of $P_k^*$  versus $K$.
We consider Rayleigh fading channel, set $K=[2,3,...,100]$, and generate 100,000 realizations of the channel vector ${\bf g}$ for each value of $K$. Since it is hard to find the same value of $g_k$ in different realizations of ${\bf g}$, we calculate the statistics of $P_k^*$ when $g_k \in (g_k'-\delta,g_k'+\delta)$, where $\delta=0.1$. Since the results are similar for different values of $g_k'$, we set $g_k'=1$. We can see that the standard deviation, which is caused by  $g_j, j \neq k$, approaches zero  when $K \geq 30$.

\subsubsection{Size Generalization by Selecting Activation Function}\label{simply-not-AF}
In what follows, we show that a mean-GNN with proper activation function at the output layer can generalize to different values of $K$ when learning the optimal power allocation policy. Denote this GNN as $\mathcal{P}_P({\bm g}; {\bm \theta}_P)$, where ${\bm \theta}_P$ is the model parameters. The output of this GNN is ${\bf {\hat p}}=[\hat{P}_1,...,\hat{P}_K]$.

From the composite function form of ${\bf F}'_3({\bf g},K)$ in \eqref{P-case2-equi}, the neural network $\mathcal{P}_P({\bm g}; {\bm \theta}_P)$ should consist of hidden layers to learn an inner function (i.e., the mapping from $g_k$ to $\hat P'_k$) and an output layer to learn the outer function (i.e., the mapping from $\hat P'_k$ to $ \hat{P}_k$). When the \texttt{softmax} function is used at the output layer,\footnote{From another composite function form of ${\bf F}'_3({\bf g},K)$,  $\mathcal{P}_P({\bm g}; {\bm \theta}_P)$ can also first learn a size-invariant inner function $f_{e1}(g_k)\triangleq \frac{1}{g_k}$and then capture the trend of the power allocation policy with $K$ by choosing other function for normalization. We choose \texttt{softmax} function as an example, since it is often used as the activation function of the output layer of DNN.} the output of $\mathcal{P}_P({\bm g}; {\bm \theta}_P)$ can be expressed as
\begin{align}
\hat{\bf p}  &= P_{\rm max} \cdot  \Big[  \frac{  e^{\hat{P}_1'}}{\sum_{k=1}^K e^{\hat{P}_k'}}, \cdots,  \frac{ e^{\hat{P}_K'}}{\sum_{k=1}^K e^{\hat{P}_k'}} \Big] \nonumber \\ & =P_{\rm max} \cdot \Big [ \sigma_s (\hat{P}_1',\{\hat{P}_j'\}_{j\neq 1}), \cdots,  \sigma_s (\hat{P}_K',\{\hat{P}_j'\}_{j\neq K})\Big] =P_{\rm max} \cdot  {\bm \sigma}_s(\hat{P}_1',\cdots,\hat{P}_K'), \nonumber
\end{align}
where $\hat{P}_k'=f_{e2}(g_k)= \ln \frac{1}{g_k}$.
When $K \to \infty$, the $k$th output of $\mathcal{P}_P({\bm g}; {\bm \theta}_P)$ becomes
\begin{subequations}
\begin{align}
\hat{P}_k = \frac{ P_{\rm max}e^{\hat{P}_k'}}{\sum_{k=1}^K e^{\hat{P}_k'}} =\frac{1}{K}   \frac{ P_{\rm max}e^{\hat{P}_k'}}{\sum_{k=1}^K e^{\hat{P}_k'}/K} \to  \frac{1}{K}   \frac{ P_{\rm max}e^{\hat{P}_k'}}{ \mathbb{E}(e^{\hat{P}_k'})}, \nonumber
\end{align}
\end{subequations}
which scales with $K$ in the same trend as $P_k^*$ shown in \eqref{P-case2-3}.

\begin{figure}[!htbp]
    \centering
    \includegraphics[width=10cm,height=3.5cm]{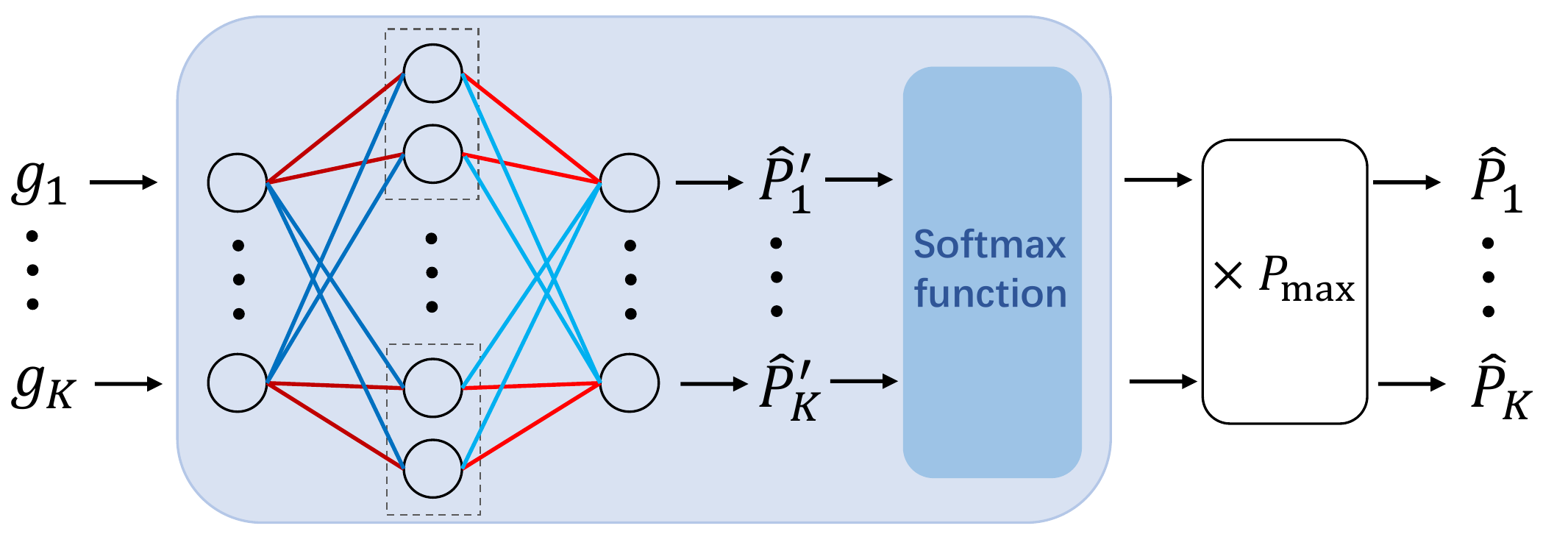}\vspace{-2mm}
    \caption{ Structure of $\mathcal{P}_P({\bm g}; {\bm \theta}_P)$. The connections with the same color are with same weights (i.e., ${\bf U}^{l}$ and ${\bf V}^{l}$ in \eqref{Out-hidd}).}
    \label{GNN-P-struc}\vspace{-6mm}
\end{figure}

The structure of $\mathcal{P}_P({\bm g}; {\bm \theta}_P)$ is shown in Fig. \ref{GNN-P-struc}, which learns a  size-invariant inner function $f_{e2}(g_k)$ before the activation function in the output layer (i.e., , $\hat{P}_k' \approx f_{e_2}(g_k)$) and then captures the asymptotic trend of the power allocation policy with $K$ (i.e., $1/K$) by \texttt{softmax} function.

\vspace{-2mm}

\vspace{-2mm}\begin{rem}
In  order to make $\hat{P}_k \approx P_k^*$, $\hat{P}_k'$ only needs to satisfy $\hat{P}_k'\approx\ln \frac{1}{g_k}+\zeta(K)$, where $\zeta(K)$ denotes any function of $K$. This is because when $\hat{P}_k'\approx\ln \frac{1}{g_k}+\zeta(K)$, we have
\begin{align}
\hat{P}_k & \approx \frac{ P_{\rm max}e^{\ln\frac{1}{g_k}+\zeta(K)}}{\sum_{k=1}^K e^{\ln\frac{1}{g_k}+\zeta(K)} } =  \frac{ P_{\rm max}e^{\ln\frac{1}{g_k}}e^{\zeta(K)} }{e^{\zeta(K)}\sum_{k=1}^K e^{\ln\frac{1}{g_k}}} = \frac{ P_{\rm max}e^{\ln \frac{1}{g_k}}}{\sum_{k=1}^K e^{\ln\frac{1}{g_k}}} = P_k^*. \nonumber
\end{align}
It implies that the GNN used to learn the inner function does not necessarily be invariant to $K$. This may explain why the power allocation policy can be learned with size generalization ability by a GNN using sum-aggregator and \texttt{softmax} in the output layer in \cite{guo2021learning}
\end{rem}

Since $B^*$ has to be found from a transcendental equation in \eqref{P-case2-1b}, $f'_4({\bf g},K)$ has no closed-form and can not be expressed as a composition of outer and inner functions explicitly. Hence, it cannot be learned by mean-GNN with size generalization ability via selecting an activation function or normalization. We will show how to deal with this issue in the next subsection.

\vspace{-4mm}
\subsection{Power and Bandwidth Allocation Policy for URLLC}
Consider a downlink orthogonal frequency division multiple access system supporting URLLC, where a BS equipped with $N_t$ antennas serves $K$ single-antenna users. To improve the resource usage efficiency, the power and bandwidth allocation is optimized to minimize the total bandwidth required to satisfy the QoS of every user with maximal transmit power $P_{\rm max}$ at the BS as,
\vspace{-2mm}
\begin{align}
\min_{P_k,B_k} \ \ & \sum_{k=1}^K B_k  \label{P1}\\
s.t. \  \ &C_k^E \geq S^E,  \tag{\ref{P1}a} \label{P1_a} \\
&  \sum_{k=1}^K P_k \leq P_{\rm max}, \tag{\ref{P1}b} \label{P1_b}\\
& P_k \geq 0, B_k\geq 0,  \tag{\ref{P1}c} \label{P1_c}
\end{align}
where $P_k$ and $B_k$ are respectively the power and bandwidth allocated to the $k$th user, $C_k^E$ is the effective capacity reflecting the system service ability for a user with QoS exponent $\theta$ depending on the queueing delay bound and delay bound violation probability \cite{EC,JT2007augTWC}, and $S^E$ is the effective bandwidth depending on the statistics of packet arrival \cite{Frank1996Notes,LL2007TIT}. \eqref{P1_a} is the QoS requirement of each user \cite{SCJ2019PIMRC}, and \eqref{P1_b} is the total transmit power constraint.

To ensure the ultra-low transmission delay, we consider a short frame structure as in  \cite{SCJ2019PIMRC}, where each frame is with duration $T_f$ and consists of a duration for data transmission and a duration for signalling. The duration for data transmission in each frame is $\tau$. Since the delay bound
is typically shorter than the channel coherence time, time diversity cannot
be exploited. To guarantee the transmission reliability
within the delay bound, we assign each user with different subcarriers in
adjacent frames   \cite{SCJ2019PIMRC}. When the frequency interval between adjacent
subcarriers is larger than the coherence bandwidth,
the small scale channel gains of a user among frames are
independent. Effective capacity can be expressed as $C_k^E= -\frac{1}{\theta } \ln \mathbb{E}_{g_k} \{ e^{-\theta s_k } \}$ packets/frame \cite{SCJ2019PIMRC}, and $s_k$ is the achievable rate of the $k$th user that can be approximated as \cite{WYang2014}
\begin{align}
s_k \approx \frac{\tau B_k}{ \mu \ln 2} \left[  \ln   \left(  1  +  \frac{\alpha_k P_k g_k }{N_0B_k}\right)  -  \frac{Q_G^{-1}(\epsilon_k^c)}{ \sqrt{\tau B_k } } \right], \label{sk-URLLC}
\end{align}
where $\mu$ is the packet size, $\alpha_k$ and $g_k$ are respectively the large-scale channel gain and two norm of the channel vector of the $k$th user, $\epsilon_k^c$ is the decoding error probability, and $Q_G^{-1}(\cdot)$ is the inverse of the Gaussian Q-function.

From the KKT conditions, the optimal solution of the problem in \eqref{P1} should satisfy
\begin{subequations}
\begin{align}
 & \sum_{k=1}^K P_k^* =P_{\rm max}, \label{Case3-sumP}\\ & C_k^E=-\frac{1}{\theta } \ln \mathbb{E}_{g_k} \{ e^{-\theta s_k(g_k,\alpha_k,P_k^*,B_k^*) } \}=S^E,\label{Case3-QoS}
\end{align}
\end{subequations}
where $s_k(g_k,\alpha_k,P_k^*,B_k^*)$ denotes the rate achieved with $P_k^*$ and $B_k^*$ given channels $\alpha_k$ and $g_k$.

Since the effective capacity is hard to derive, we can not obtain closed-form solution of problem \eqref{P1}. When channel distribution is unknown, the solution cannot be obtained even with numerical algorithms. Therefore, we resort to DNNs to learn the mapping from environment parameters to $P_k^*$ and $B_k^*$. We can see from \eqref{Case3-QoS} that the effective capacity of the $k$th user depends on $\alpha_k$, and hence the environment parameters are the large scale channel gains.

Denote the optimal power allocation policy as ${\bf p}^*={\bf P}'({\bm \alpha},K)$ and the bandwidth allocation policy as ${\bf B}^*={\bf B}'({\bm \alpha},K)$, where ${\bf p}^*=[P^*_{1},...,P^*_{K}]$, ${\bf B}^*=[B^*_{1},...,B^*_{K}]$, and  ${\bm \alpha}=[\alpha_{1},...,\alpha_{K}]$. Since the multivariate functions ${\bf P}'({\bm \alpha},K)$ and ${\bf B}'({\bm \alpha},K)$ are not affected by the order of the users, both are PE policies and hence can be respectively learned by GNNs.

We use mean-GNNs for learning ${\bf P}'({\bm \alpha},K)$ and ${\bf B}'({\bm \alpha},K)$, which are denoted as $\mathcal{P}_P({\bm \alpha};{\bm \theta}_P)$ and $\mathcal{P}_B({\bm \alpha};{\bm \theta}_B)$, respectively, where ${\bm \theta}_P $ and ${\bm \theta}_B $ are the model parameters. The input of both GNNs is ${\bm \alpha}$.  The output of  $\mathcal{P}_P({\bm \alpha};{\bm \theta}_P)$ is $[\hat{P}_1,...,\hat{P}_K]$, and the output of $\mathcal{P}_B({\bm \alpha};{\bm \theta}_B)$ is $ [\hat{B}_1,...,\hat{B}_K]$.

Due to the total power constraint shown in \eqref{Case3-sumP}, $P_k^*$ decreases with $K$. As a result, $B_k^*$ should increase with $K$ to ensure the QoS in \eqref{Case3-QoS}. This indicates that the two policies do not satisfy the size-invariance property.

\subsubsection{Size Generalization by Pre-training ``Activation Function''}
${\bf P}'({\bm \alpha},K)$ cannot be expressed as a composite function in closed-form as in \eqref{P-case2-equi}, but it is not size-invariant due to the same reason as ${\bf F}'_3({\bf g},K)$: the summation constraint, which is also a kind of prior knowledge. Inspired by the analysis in section \ref{simply-not-AF}, we can choose \texttt{softmax} as the activation function of the output layer of $\mathcal{P}_P({\bm \alpha}; {\bm \theta}_P)$ in order for the mean-GNN generalizable to $K$.

${\bf B}'({\bm \alpha},K)$ also cannot be expressed as a composite function in closed-form, which is not size-invariant due to the QoS constraint and its dependence with $P_k^*$.
Inspired by the structure of $\mathcal{P}_P({\bm g}; {\bm \theta}_P)$ in Fig. \ref{GNN-P-struc}, we conceive the following approach to enable mean-GNN generalizable to $K$ for learning ${\bf B}'({\bm \alpha},K)$.
We first pre-train an ``activation function'' to find the scaling law of ${\bf B}'({\bm \alpha},K)$ with $K$. Then, we use a mean-GNN with SI-AFs in hidden layers to learn a size-invariant inner function and use the pre-trained ``activation function'' in the output layer of the mean-GNN.
Such an approach is based on a conjecture that ${\bf B}'({\bm \alpha},K)$ is a composite function of an inner function independent of $K$ and an outer function dependent on $K$.

Given a power allocation policy, ${\bf B}'({\bm \alpha},K)$ can be degenerated into a function of $\alpha_k$ and $K$ when $K \rightarrow \infty$ as shown in \eqref{Define_fk}
i.e., $B_k^* = B^v(\alpha_k,K)$. This function reflects the scaling law of bandwidth allocation policy with $K$ and is identical for all users, which can approximate the implicit outer function of the PE policy in the non-asymptotic regime.

\begin{figure}[!htbp]
    \centering
    \includegraphics[width=8cm,height=3cm]{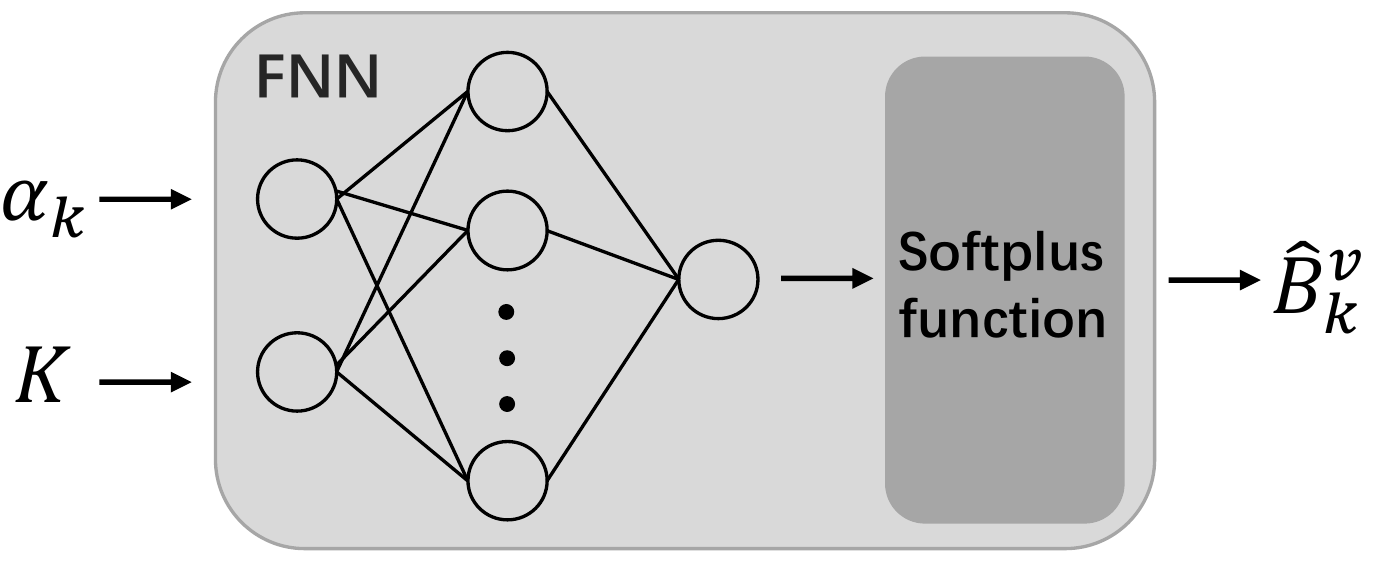}\vspace{-3mm}
    \caption{Structure of $\mathcal{B}^v(\alpha_k,K;{\bm \theta}_v)$ for learning ${B}^{v}(\alpha_k,K)$.}
    \label{FCNN-struc}
\end{figure}

In order to find the function ${B}^{v}(\alpha_k,K)$ for all possible values of $\alpha_k$ and $K$, we introduce a FNN denoted as $\mathcal{B}^v(\alpha_k,K;{\bm \theta}_v)$, where ${\bm \theta}_v$ is the model parameters. The structure of $\mathcal{B}^v(\alpha_k,K;{\bm \theta}_v)$ is shown in Fig. \ref{FCNN-struc}, where the input is $[\alpha_k,K]$, and the output is $\hat{B}^v_k$.
\vspace{-2mm}
\vspace{-2mm}

To reduce the cost for training, we pre-train $\mathcal{B}^v(\alpha_k,K;{\bm \theta}_v)$ and assume equal power allocation during the pre-training. The samples for $\alpha_k$ are generated from random realizations of large scale channel and the samples for $K$ are all possible numbers of users in the considered system.
The FNN can be pre-trained in supervised manner, where the model parameters are found from
\begin{align}
 {\bm \theta}_v =\arg \min_{{\bm \theta}_v} &\mathbb{E}_{\alpha_k,K} \big\{  (\hat{B}^v(\alpha_k,K;{\bm \theta}_v)-B^{v*}  )^2 \big\}, \label{Loss-sup}
\end{align}
where $\hat{B}^v(\alpha_k,K;{\bm \theta}_v)$ denotes the function learned by $\mathcal{B}^v(\alpha_k,K;{\bm \theta}_v)$, and $B^{v*}$ is the label obtained by numerically solving the equation in \eqref{Case3-QoS} with bi-section method by setting $P_k^*=\frac{P_{\rm \max}}{K}$ and estimating the expectation as empirical mean from samples.
The FNN can also be pre-trained in unsupervised manner, where the model parameters are found from
\begin{align}
 {\bm \theta}_v =\arg \min_{{\bm \theta}_v} \mathbb{E}_{\alpha_k,K} \Big\{  \big( S^E+   \frac{1}{\theta }  \ln \mathbb{E}_{{g_k} }  \{ e^{-\theta s_k(g_k,\alpha_k,\frac{P_{\rm max}}{K},\hat{B}^v(\alpha_k,K;{\bm \theta}_v))} \}\big)^2 \Big\}, \label{Loss-unsup}
\end{align}
where the loss function is from the KKT condition in \eqref{Case3-QoS}.

\begin{figure}[!htbp]
    \centering
    \includegraphics[width=10cm,height=3.5cm]{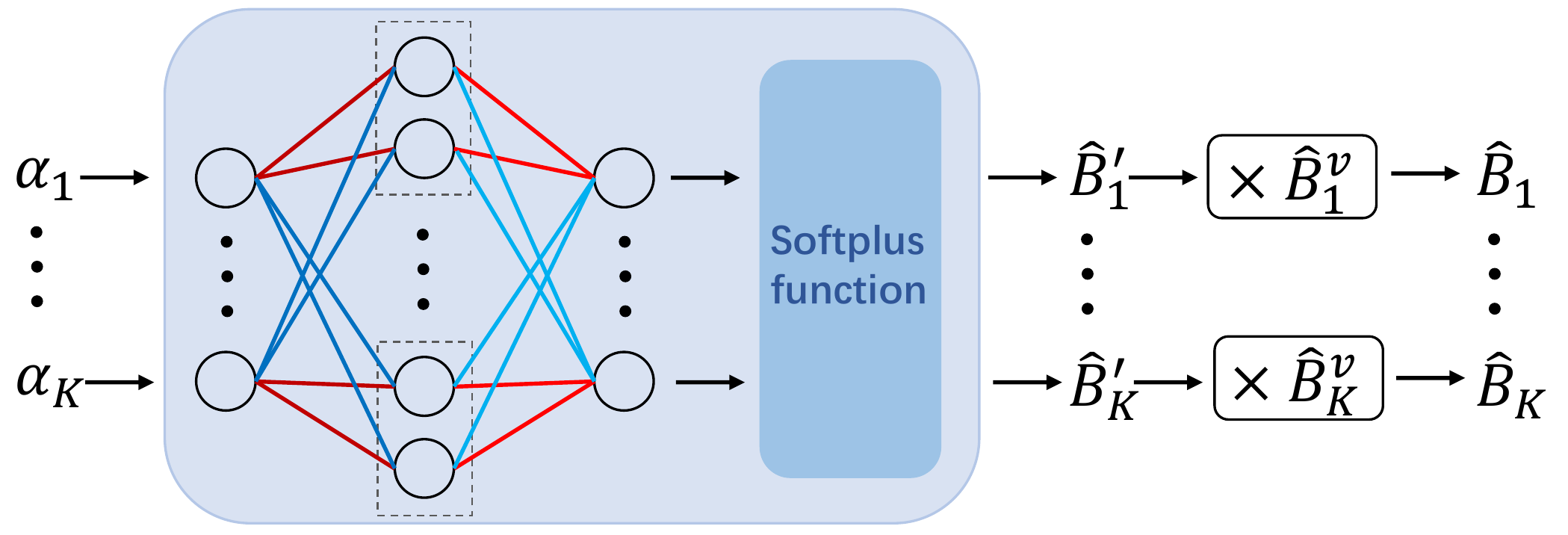}\vspace{-4mm}
    \caption{ Structure of $\mathcal{P}_B({\bm \alpha}; {\bm \theta}_B)$ for learning ${\bf B}({\bm \alpha})$.}
    \label{GNN-B-struc}\vspace{-6mm}
\end{figure}

The structure of $\mathcal{P}_B({\bm \alpha}; {\bm \theta}_B)$ for learning ${\bf B}'({\bm \alpha},K)$ with size generalization ability is shown in Fig. \ref{GNN-B-struc}. It composes a mean-GNN with SI-AFs in hidden layers for learning the size-invariant inner function (i.e., the mapping from $\alpha_k$ to $\hat B_k'$) and the pre-trained FNN in the output layer for learning the outer function (i.e., the mapping from $\hat B_k'$ to $\hat B_k$), where $\hat{B}_k \triangleq \hat{B}_k' \cdot \hat{B}^v_k$.
To ensure $\hat B_k \geq 0$, \texttt{Softplus} is used at the output layer of  $\mathcal{B}^v$, before the pre-trained activation function of the mean-GNN.

It is worthy to note that the prior knowledge for the bandwidth allocation problem (i.e., the optimal solution should satisfy the KKT condition) has been embedded in the pre-trained FNN, which serves as the inductive bias for aligning the size-scaling law of $\mathcal{P}_B({\bm \alpha}; {\bm \theta}_B)$ to the size-scaling law of the policy.



\vspace{-2mm}


\subsubsection{Unsupervised Learning for the Policies}
Since the closed-form expression of the effective capacity is unable to be derived in general cases, the labels for supervised learning can not be obtained by solving problem \eqref{P1}. Moreover, the constraints can not be ensured to be satisfied by supervised learning.
To control the stringent constraint in URLLC, we train the GNNs in unsupervised manner \cite{SCJ2019PIMRC}.
To this end, we first transform \eqref{P1} into its primal-dual problem as
\vspace{-2mm}
\begin{align}
 \max_{\lambda_k}\min_{P_k,B_k} \ \ & L({\bm \alpha},{P}_k,{B}_k,{\lambda}_k ) \triangleq \sum_{k=1}^K \Big[B_k+ \lambda_k( \mathbb{E}_{g_k} \{ e^{-\theta s_k}\}-e^{-\theta  S^E} ) \Big] \label{P2}\\
s.t. \  \  & \sum_{k=1}^K P_k \leq P_{\rm max}, \tag{\ref{P2}a} \label{P2_a} \\
\ \ & P_k \geq 0, B_k\geq 0, \lambda_k \geq 0, \tag{\ref{P2}b} \label{P2_b}
\end{align}
where  $ L({\bm \alpha},{P}_k,{B}_k,{\lambda}_k ) $ is the Lagrangian function of problem \eqref{P1}, and $ \lambda_k$ is the Lagrangian multiplier for the QoS constraint.

According to the proof in \cite{SCJ2019PIMRC}, the variable optimization problem in \eqref{P2} can be equivalently transformed into the following functional optimization problem,
\begin{align}
 \max_{\lambda_k({\bm \alpha})}\min_{P'_k({\bm \alpha},K),B'_k({\bm \alpha},K)}& \mathbb{E}_{\bm \alpha} \{L({\bm \alpha},P'_k({\bm \alpha},K),B'_k({\bm \alpha},K),\lambda_k({\bm \alpha}) )  \}   \label{P3}\\
s.t. \  &  \eqref{P2_a}, \eqref{P2_b}. \nonumber
\end{align}

Since the optimal Lagrangian multiplier function is not affected by the order of users, we can use a GNN $\mathcal{P}_\lambda({\bm \alpha};{\bm \theta}_\lambda)$ to approximate $\lambda_k^*({\bm \alpha})$,  where ${\bm \alpha}$ is the input and ${\bm \theta}_{\lambda}$ is the model parameters. Denote the output as $[\hat{\lambda}_1,...,\hat{\lambda}_K]$. By respectively replacing ${P'_k}(\bm \alpha,K)$, ${B'_k}(\bm \alpha,K)$ and ${\lambda_k}(\bm \alpha)$ in \eqref{P3} by the functions $\hat{P}'_k({\bm \alpha},K;{\bm \theta}_P)$, $\hat{B}'_k({\bm \alpha},K;{\bm \theta}_B)$ and $\hat{\lambda}_k({\bm \alpha};{\bm \theta}_{\lambda})$ learned by three GNNs, the optimal model parameters of the GNNs can be found from the following problem,
\begin{align}
  \max_{{\bm \theta}_{\lambda} } \min_{ { {\bm \theta}_P,{\bm \theta}_B }  } & \mathbb{E}_{\bm \alpha} \Big\{  L\big({\bm \alpha},\hat{P}_k({\bm \alpha};{\bm \theta}_P), \hat{B}_k({\bm \alpha};{\bm \theta}_B ),  \hat{\lambda}_k({\bm \alpha};{\bm \theta}_{\lambda} ) \big)  \! \Big\}   \label{P4}\\
s.t. \  \  &   \eqref{P2_a}, \eqref{P2_b}. \nonumber
\end{align}
The constraint in \eqref{P2_a} can be satisfied by selecting \texttt{softmax} as the activation function in the output layer of $\mathcal{P}_P({\bm \alpha}; {\bm \theta}_P)$, and the constraint in \eqref{P2_b} can be satisfied by selecting \texttt{softplus} as the activation function in the output layer of $\mathcal{P}_B({\bm \alpha}; {\bm \theta}_B)$, $\mathcal{B}^v(\alpha_k,K;{\bm \theta}_v)$ and $\mathcal{P}_\lambda({\bm \alpha};{\bm \theta}_\lambda)$.

The joint training procedure for $\mathcal{P}_P({\bm \alpha}; {\bm \theta}_P)$,  $\mathcal{P}_B({\bm \alpha}; {\bm \theta}_B)$ and $\mathcal{P}_\lambda({\bm \alpha}; {\bm \theta}_\lambda)$ is summarized in Algorithm \ref{Ag1}, where $t$ denotes the number of iterations, ${\bm g} \triangleq [g_1,...,g_K]$, $\delta_t$ is the learning rate, $|\cdot|$ denotes cardinality, $\nabla_{\bf x} {\bf y}=[\nabla_{\bf x} y_1,...,\nabla_{\bf x}y_K]$, $\nabla_{\bf x}y_i=[\frac{\partial y_i}{\partial x_1},...,\frac{\partial y_i}{\partial x_K}]^T$ and $L^t({\bm g},{\bm \alpha},\hat{P}_k,\hat{B}_k,\hat{\lambda}_k ) \triangleq  \sum_{k=1}^K \Big[\hat{B}_k+$ $\hat{\lambda}_k(   e^{-\theta s_k}  -e^{-\theta S^E} ) \Big]$.

\begin{algorithm}[h]
  \caption{Training Procedure.}
  \label{Ag1}
  \begin{algorithmic}[1]
    \State Initialize  $\mathcal{P}_P({\bm \alpha}; {\bm \theta}_P)$,  $\mathcal{P}_B({\bm \alpha}; {\bm \theta}_B)$ and $\mathcal{P}_\lambda({\bm \alpha}; {\bm \theta}_\lambda)$ with random parameters ${\bm \theta}_P$, ${\bm \theta}_B$ and ${\bm \theta}_\lambda$.
    \For{ $t = 1,2,...$}
      \State Sample a batch of $({\bm g}, {\bm \alpha})$ as $\mathcal{S}$.
      \State Input ${\bm \alpha}$ into $\mathcal{P}_P({\bm \alpha}; {\bm \theta}_P)$,  $\mathcal{P}_B({\bm \alpha}; {\bm \theta}_B)$ and $\mathcal{P}_\lambda({\bm \alpha}; {\bm \theta}_\lambda)$ to obtain  $\hat{P}_k$, $\hat{B}_k'$ and $\hat{\lambda}_k$, respectively.
      \For{ $k = 1,...,K$}
         \State Input $[\alpha_k,K]$ into the pre-trained $\mathcal{B}^v(\alpha_k,K;{\bm \theta}_v)$ to obtain $\hat{B}_k^v$. \label{Bv-fd}
         \State Scale $\hat{B}_k'$  with $\hat{B}_k^v$ to obtain $\hat{B}_k $, i.e., $\hat{B}_k =\hat{B}_k' \cdot \hat{B}^v_k$. \label{Bv-getB}
      \EndFor
      \State Update the parameters  ${\bm \theta}_P$, ${\bm \theta}_B$ and ${\bm \theta}_\lambda$ via stochastic gradient descend method as follows,
      \vspace{-3mm}
      \begin{subequations}
        \begin{align}
        \ \ {\bm \theta}_{{P}}^{t+1}  \!\leftarrow\! {\bm \theta}_{{P}}^{t} - &\frac{\delta_t}{|\mathcal{S}|}\! \sum\limits_{({\bm g},{\bm \alpha}) \in \mathcal{S}} \! \nabla_{ {\bm \theta}_{P}} \mathcal{ P}_P({\bm \alpha};{\bm \theta}_{P}) \frac{\partial L^t({\bm g},{\bm \alpha},\hat{P}_k,\hat{B}_k,\hat{\lambda}_k )}{\partial \hat{P}_k},    \nonumber \\
        \ \ {\bm \theta}_{{B}}^{t+1}  \!\leftarrow \!{\bm \theta}_{{B}}^{t} - &\frac{\delta_t}{|\mathcal{S}|} \! \sum\limits_{({\bm g},{\bm \alpha}) \in \mathcal{S}} \! \nabla_{ {\bm \theta}_{B}} \mathcal{P}_B({\bm \alpha};{\bm \theta}_{B}) \frac{\partial L^t({\bm g},{\bm \alpha},\hat{P}_k,\hat{B}_k,\hat{\lambda}_k )}{\partial \hat{B}_k} \hat{B}_k^v, \nonumber \\
        \ \ {\bm \theta}_{{\lambda}}^{t+1}  \!\leftarrow \!{\bm \theta}_{{\lambda}}^{t} + & \frac{\delta_t}{ |\mathcal{S}| } \!\sum\limits_{({\bm g},{\bm \alpha}) \in \mathcal{S}} \! \nabla_{ {\bm \theta}_{\lambda}} {\mathcal{P}_{\lambda}}({\bm \alpha};{\bm \theta}_{\lambda}) \frac{\partial L^t({\bm g},{\bm \alpha},\hat{P}_k,\hat{B}_k,\hat{\lambda}_k )}{\partial \hat{\lambda_k}}. \nonumber
        \end{align}
      \end{subequations}\vspace{-0.2mm}
\EndFor
\end{algorithmic}\vspace{-0.2mm}
\end{algorithm}

\vspace{-2mm}
\section{Simulation Results}
In this section, we evaluate the size generalization performance of the proposed mean-GNNs for learning the power and bandwidth allocation in URLLC.

\vspace{-2mm}
\subsection{System Setup}
The cell radius is 250m. All users are randomly located in a road with the minimal distance of 50 m from the BS. The path loss model is $35.3+37.6\log(d_k)$, where $d_k$ is the distance between the $k$th user and the BS. The maximal transmit power of the BS and the number of antennas are respectively set as $P_{\rm \max}=43$ dBm and $N_t=8$. The single-sided noise spectral density is $N_0=-173$ dbm/Hz.
The required overall packet loss probability  and the delay bound are respectively $\epsilon_{\rm \max}=10^{-5}$  and $D_{\rm max}=1$ ms. After subtracting the downlink transmission delay and the decoding delay from the delay bound, the queueing delay is bounded by $0.8$ ms \cite{SCJ2019PIMRC}. As in \cite{SCY2018}, the decoding error probability $\epsilon_k^c$ and the queueing delay violation probability $\epsilon_k^q$ are set as $\epsilon_k^c=\epsilon_k^q= \epsilon_{\rm \max}/2$. The time for data transmission and the frame duration are $\tau=0.05$ ms and $T_f=0.1$ ms, respectively.  The packet size is $\mu=20$ bytes \cite{3GPP_rate}. The packets arrive at the BS according to Poisson process with average arrival rate $a=0.2$ packets/frame. Then, $S^E=\frac{a}{\theta }(e^{\theta }-1)$ packets/frame and $\theta =\ln[1-\frac{\ln(\epsilon_{\rm \max}/2)}{a D_{\rm \max}/{T_f} } ]$\cite{SCY2018}.

When the system parameters such as $P_{\rm \max}$, $N_t$, $\tau$, $T_f$, $\epsilon_{\rm \max}$, $D_{\rm max}$ and user locations change, similar results can be obtained and hence are not provided for conciseness.

\vspace{-4mm}\subsection{Sample Generation and Fine-tuned Hyper-parameters}
For notational simplicity, denote the three GNNs as $\mathcal{P}_P({\bm \alpha};{\bm \theta}_P) \triangleq \mathcal{P}_P$, $\mathcal{P}_B({\bm \alpha};{\bm \theta}_B) \triangleq \mathcal{P}_B$  $\mathcal{P}_\lambda({\bm \alpha};{\bm \theta}_\lambda) \triangleq \mathcal{P}_\lambda$ and the FNN as $\mathcal{B}^v(\alpha_k,K;{\bm \theta}_v) \triangleq \mathcal{B}^v$  in the sequel.

Since the loss function in \eqref{Loss-sup} is simpler than \eqref{Loss-unsup} and the label for pre-training $\mathcal{B}^v $ can be easily obtained by solving \eqref{Case3-QoS} with training samples, we use supervised learning to train this FNN.
Denote one sample for pre-training $\mathcal{B}^v$ as $[(\alpha_k,K),B^{v*}]$, where $\alpha_k$ is obtained from the path loss model with randomly generated locations, and $B^{v*}$ is the label. We generate $4,000$ samples to pre-train the FNN, where $20$ samples are generated for every $K$ from $1$ to $200$.

Denote one sample for training $\mathcal{P}_P$, $\mathcal{P}_B$, and $\mathcal{P}_\lambda$ as $({\bm g}, {\bm \alpha})$, which is a realization of the small and large-scale channel gains of all users. In each realization, ${\bm g}$ is randomly generated from Rayleigh distribution and ${\bm \alpha}$ is obtained from the path loss model with randomly generated locations of all users.
The training, validation and test sets are generated according to Table. \ref{Tr-Te-sets}.
In particular, $2000$ samples are generated for $K=10$ (i.e., all these samples for the vectors ${\bm g}$ and ${\bm \alpha}$ are of size 10) when generating training set, $200$ samples are generated for $K=10$ when generating validation set, and $100$ samples are generated for every $K$ in $[1,2,5,10,50,100,200]$  (i.e., 100 samples for ${\bm g}$ and ${\bm \alpha}$ are scalars, 100 samples for ${\bm g}$ and ${\bm \alpha}$ are of size two, 100 samples for the vectors are of size five, \emph{etc.})  when generating test set. Therefore, the total number of samples in the training, validation and test sets are $2000$, $200$ and $700$, respectively.
These three datasets are used as follows unless otherwise specified.
\vspace{-3mm}\begin{table}[htbp]
\centering
\caption{Generating Samples in Training, validation and test sets.}
\label{Tr-Te-sets}
\vspace{-4mm}
\footnotesize
\begin{tabular}{c|c|c|c}
\hline
\hline
  &  \tabincell{c} {Sizes $K$ in samples}  &   \tabincell{c} { Number of samples \\ generated for each $K$ } & \tabincell{c} {Total number of samples}  \\
\hline
\tabincell{c}{ Training set }    & \tabincell{c} { $[10 ]$  } &  \tabincell{c} {  $2,000$    }   & $2,000$ \\
\hline
  \tabincell{c}{Validation set  }   & \tabincell{c} { $[10]$}  &   \tabincell{c} {  $200$   }   & $200$\\
 \hline
  \tabincell{c}{Test set  }    & \tabincell{c} { $[1,2,5,10,50,100,200]$}   &   \tabincell{c} {  $100$   } &  $700$\\
\hline
\hline
\end{tabular}
\end{table}
\vspace{-3mm}

The fine-tuned hyper-parameters are shown in Table. \ref{Hypara}, where $[*_1,*_2,...,*_L]$ denotes that a DNN is with  $*_l$ neurons in the $l$th hidden layer. Considering that the large-scale channel gains are usually small, we scale them as $\ln({\bm \alpha})+30$. In order to make the input variables being in the same order of magnitude when pre-training $\mathcal{B}^v$, we also scale $K$ as $K/200$.

\vspace{-2mm}\begin{table}[!htbp]
\centering
\caption{Hyper-parameters.}
\label{Hypara}
\vspace{-4mm}
\footnotesize
\begin{tabular}{c|c|c|c|c}
\hline
\hline
\multirow{2}*{ Hyper-Parameters } & \multicolumn{3}{c|}{GNN} & \multirow{2}*{\tabincell{c}{FNN $\mathcal{B}^v$}}\\
\cline{2-4}
    & $\mathcal{P}_P $  & $\mathcal{P}_B $  &  $\mathcal{P}_\lambda $  &  \\
\hline
\tabincell{c}{Number of neurons in hidden layers} & \multicolumn{3}{c|} { $[4K]$ } & $[200,100,100,50]$ \\
\hline
\tabincell{c}{Activation function of the hidden layers} & \multicolumn{4}{c} {  \tabincell{c}{\texttt{Leaky ReLU}  ( i.e.,$y=x,x\geq0, $  $ y=0.01x,x<0$ )}  }\\
\hline
\tabincell{c}{Activation function of the output layer} &  \tabincell{c}{\texttt{Softmax}}   & \multicolumn{3}{c}{  \tabincell{c}{\texttt{Softplus}    }} \\
\hline
\tabincell{c}{Learning rate} & \multicolumn{3}{c|} { $\frac{0.01}{1+0.01t}$}  & $\frac{0.01}{1+0.001t}$ \\
\hline
\tabincell{c}{Batch size} & \multicolumn{3}{c|} {$10$} & 100 \\
\hline
\tabincell{c}{Epochs} & \multicolumn{3}{c|} {$5000$} & $2500$  \\
\hline
\hline
\end{tabular}
\end{table}

\vspace{-6mm}
\subsection{Size Generalization Performance}
The system performance is measured by the availability and  total bandwidth achieved by $[\hat{P}_1,...,\hat{P}_K]$ and $ [\hat{B}_1,...,\hat{B}_K]$. Availability is the percentage of the users with satisfied QoS among all the users in the system that is one of the key performance metrics for URLLC  \cite{Ava}, and total bandwidth is the sum of the bandwidth of all users achieved by a learned policy. The two metrics are respectively defined as,
\begin{align}
{\rm A}_K & \triangleq \frac {1} {K|\mathcal{N}_K|} {\sum_{({\bm g},{\bm \alpha}) \in \mathcal{N}_K} \sum_{k=1}^K I(\epsilon_k < \epsilon_{\rm max})}, ~~
{\rm W}_K^{\rm r} \triangleq \frac{1}{ |\mathcal{N}_{K} |}\sum_{({\bm g},{\bm \alpha}) \in \mathcal{N}_K}{\sum_{k=1}^K \hat{B}_k} \ \ \ {\rm (MHz)}, \nonumber
\end{align}
where $\mathcal{N}_{K} $ is the subset in the test set containing the samples for the channel vectors with the same size, and $\epsilon_k$ is the packet loss probability of the $k$th user achieved by a learned policy.

To reflect the impact of the randomness of small training sets, we train the GNNs 10 times. In each time, the training set is randomly generated and the GNNs are randomly initialized while the test set remains unchanged. The values of the two metrics are obtained by selecting the second worst test results with 10 well-trained DNNs, hence are with confidence level of $90\%$.

\subsubsection{Reliability Controlling}
The learned solution is unable to satisfy the QoS with probability one. As demonstrated in \cite{SCJ2019PIMRC}, the packet loss probability can be ensured by setting a more stringent requirement for reliability than the required reliability $\epsilon_{\rm \max}$ during training, i.e. $\epsilon_D < \epsilon_{\rm \max}$. With such a conservative design, even if the overall packet loss probability achieved by the learnt solution exceeds $\epsilon_D$, the probability of violating the original requirement $\epsilon_{\rm \max}$ is low (e.g., the availability is $1.00$ for $\epsilon_D=6 \times 10^{-6}$ as shown in the following table).

The impacts of different values of $\epsilon_D$ on the system performance are shown in Table. \ref{SP-Tb}. We can see that by reducing $\epsilon_D$, the availability can be remarkably improved with a slight increase of the total bandwidth. For $\epsilon_D = 6\times10^{-6}$, the availability achieves $1.00$ with a bandwidth loss $\frac{53.14-51.81}{51.81} \approx 2.57\%$ from $\epsilon_D = 10^{-5}$ when $K=200$ in the test set. Recalling that $K=10$ in training set as in Table \ref{Tr-Te-sets}, the results in Table. \ref{SP-Tb} indicate that the required reliability $\epsilon_{\rm \max}$ can be ensured even when the well-trained GNNs are tested in the scenario with different number of users from the training set.


\begin{table*}[htbp]
\centering
\caption{Impacts of the values of $\epsilon_D$ on system performance.}
\label{SP-Tb}
\vspace{-4mm}
\footnotesize
\begin{tabular}{ c|c|c|c|c|c|c|c|c|c|c|c|c|c|c}
\hline
\hline
   \multirow{2}*{\tabincell{c}{ $\epsilon_D$}}  &  \multicolumn{2}{c|}{$K=1$}&  \multicolumn{2}{c|}{$K=2$} & \multicolumn{2}{c|}{$K=5$} &  \multicolumn{2}{c|}{$K=10$} &   \multicolumn{2}{c|}{$K=50$} &     \multicolumn{2}{c|}{$K=100$}   &     \multicolumn{2}{c}{$K=200$}   \\
\cline{2-15}
 &      ${\rm A}_K$ & ${\rm W}_K^{\rm r}$  &   ${\rm A}_K$ & ${\rm W}_K^{\rm r}$ &  ${\rm A}_K$ & ${\rm W}_K^{\rm r}$ &   ${\rm A}_K$ & ${\rm W}_K^{\rm r}$  &  ${\rm A}_K$ & ${\rm W}_K^{\rm r}$  &   ${\rm A}_K$ & ${\rm W}_K^{\rm r}$  &   ${\rm A}_K$ & ${\rm W}_K^{\rm r}$ \\
\hline
\tabincell{c}{ $10^{-5}$}& 0.57 & 0.13  &    0.26 & 0.27 & 0.32 & 0.77 &   0.44 & 1.66  & 0.42 & 10.29  & 0.41 & 22.95   & 0.45 & 51.81        \\
\hline
\tabincell{c}{ $8  \times  10^{-6}$}& 1.00 & 0.14 &     0.81 & 0.28 & 0.84 & 0.78 &   0.91 & 1.68   & 0.77 & 10.38  & 0.73 & 23.15   & 0.63 & 52.27         \\
\hline
\tabincell{c}{  $ {\bf 6  \times  10^{-6} }$  }& {\bf 1.00} & {\bf 0.14} &    {\bf 1.00}&  {\bf 0.28}& {\bf 1.00}& {\bf 0.79} &   {\bf 1.00} & {\bf 1.70}   & {\bf 1.00} & {\bf 10.55}  & {\bf 1.00}& {\bf 23.53}   & {\bf 1.00} & {\bf 53.14}        \\
\hline
\tabincell{c}{ $5  \times  10^{-6}$}& 1.00 & 0.14   &      1.00 & 0.28 & 1.00 & 0.80 &   1.00 & 1.72 & 1.00 & 10.64  & 1.00& 23.74   & 1.00 & 53.60       \\
\hline
\hline
\end{tabular}\vspace{-3mm}
\end{table*}

\vspace{-4mm}\subsubsection{Impact of Small $K$}
In previous analytical analysis, $K$ is assumed approaching infinity. In what follows, we reduce the value of $K$ in training set to show that the proposed method can still be applied for small number of users. In Table. \ref{NumerSamp-diffK}, we show the minimal number of samples and the time required to train the GNNs  (i.e., sample complexity and time complexity) when the training samples are with different sizes for achieving the availability of $1.00$ in all test scenarios, where the test samples are generated with size of $1, 2, 5, 10, 50, 100$ or 200 in each scenario. It is shown that the learned policies by the proposed GNNs can achieve good generalization performance for large size (e.g., $K=200$) even trained with samples of small size (e.g., $K=2$), despite that the number of training samples and training time increase with the decreasing sizes of training samples. The increased training complexity comes from a fact that the output of the pre-trained FNN, $\mathcal{B}_v$, is not an accurate approximate of ${B}_k^* $ 
 when $K$ is small, which calls for more training samples to make $\mathcal{P}_B$ to learn the policy accurately.

\begin{table}[htbp]
\centering
\caption{Training complexity for achieving the availability of $1.00$,  $\epsilon_D=6\times10^{-6}$.}
\vspace{-4mm}
\footnotesize
\label{NumerSamp-diffK}
\begin{tabular}{c|c|c}
\hline
\hline
   \tabincell{c} {Size $K$ of training samples}  &   \tabincell{c} { Number of training samples  } &   \tabincell{c} { Training time (s) }    \\
\hline
 \tabincell{c} { [$2$]  } &  \tabincell{c} {  $5,000$    } &  1,290.32  \\
\hline
\tabincell{c} { [$5$]}  &   \tabincell{c} {  $4,000$   }  & 1,289.91 \\
 \hline
\tabincell{c} { [$8$]}   &   \tabincell{c} {  $3,000$   }  & 1,185.11\\
\hline
\tabincell{c} { [$10$]}   &   \tabincell{c} {  $2,000$   } & 875.43 \\
\hline
\hline
\end{tabular}\vspace{-3mm}
\end{table}

\vspace{-2mm}\subsubsection{Performance Comparison}
To evaluate the performance of the proposed GNN in terms of size generalization, we compare the system performance of the following three DNNs.
\begin{itemize}
	\item ``P-GNN": This is the proposed GNNs, where the power allocation policy is learned by $\mathcal{P}_P$ with \texttt{softmax} in output layer and the bandwidth allocation policy is learned by $\mathcal{P}_B $ with the pre-trained ``activation function" at the output layer, i.e., $\hat{B}_k  = \hat{B}_k' \hat{B}^v_k$.
	\item  ``M-GNN": This is existing mean-GNNs, where the power allocation policy is learned by $\mathcal{P}_P $ with \texttt{softmax} in output layer, but the bandwidth allocation policy is learned by a mean-GNN with SI-AFs in all layers (which is $\mathcal{P}_B $ without pre-trained ``activation function" at the output layer, i.e., $\hat{B}_k = \hat{B}_k' $).
\item  ``FNN":  The power and bandwidth allocation policies are learned by two FNNs. Each FNN is with one hidden layer consisting of 300 neurons. Both the input size and output size are $K_{\rm max}$. When the training or test samples are generated with $K < K_{\rm max}$, the input vector is padded with zeros. The learning rate is $\frac{0.001}{1+0.001t}$, the batch size is $100$ and the other hyper-parameters are the same as that of the GNNs in Table \ref{Hypara}.
\end{itemize}

We first provide the minimal number of training samples and the values of $\epsilon_D$ required by each DNN to converge in Table. \ref{NumerSamp-diffmeth}, where both the availability and total bandwidth can be improved less than $1\%$ by further increasing the training samples and epoches. We also provide the training time to reflect the time complexity of each DNN.
Since ``FNN" can only be with fixed-size inputs, we use the samples with $K_{\rm max} =200$ to train the FNN such that it can be used for the test samples with variable sizes of $K \leq 200$, where the test samples with $K<200$ are padded with zeros. For ``M-GNN", we respectively use the training samples of sizes $K=10$ and $K=200$ for comparison. Since $\epsilon_D = 6 \times 10^{-6}$ is unable to control the reliability achieved by the policies learned by ``M-GNN" and ``FNN", the values of $\epsilon_D $ chosen for different DNNs are also provided in Table. \ref{NumerSamp-diffmeth}.

\vspace{-2mm}\begin{table}[htbp]
\centering
\caption{Minimal number of samples for training and training time.}
\vspace{-4mm}
\footnotesize
\label{NumerSamp-diffmeth}
\begin{tabular}{c|c|c|c}
\hline
\hline
 Methods ($\epsilon_D$) & \tabincell{c} {$K$ in training samples}   & \tabincell{c} {Training samples}  & Training time (s) \\
\hline
 P-GNN ($6 \times 10^{-6}$) & \tabincell{c} { [$10$]  }   & $2,000$ & 875.43  \\
\hline
 M-GNN ($5 \times 10^{-6}$)&\tabincell{c} { [$10$]}    & $2,000$  &  503.81\\
 \hline
 M-GNN ($3 \times 10^{-6}$) &\tabincell{c} { [$200$]}    & $300$ &  429.71  \\
\hline
 FNN ($2 \times 10^{-6}$) &\tabincell{c} { [$200$]}   &  $100,000$  & 49,422.77\\
\hline
\hline
\end{tabular}\vspace{-3mm}
\end{table}

The system performance achieved by the learned policies in the  test scenarios where the test samples are generated with size of 1, 2, 5, 10, 50, 100 or 200 in each scenario is shown in Table. \ref{SP-diffGNN}, where ``M-GNN$^{K}$" denotes the  ``M-GNN" trained by the samples with size of $K$. The number of samples used for training and the corresponding values of $\epsilon_D$ used in training are shown in Table. \ref{NumerSamp-diffmeth}. Hence, ``M-GNN$^{200}$" and  ``FNN" are trained solely on the maximal-sized samples.
It is shown that when $K=10$ and $K=200$ in the test set where ``M-GNN$^{10}$" and ``M-GNN$^{200}$"  do not need to generalize, their availability is ${1.00}$ and the required total bandwidth is close to ``P-GNN".
The availability of ``M-GNN$^{10}$"  is ${0.00}$ in the test scenarios where $K>10$, i.e., the QoS of all users are not satisfied. Although the availability of ``M-GNN$^{10}$"  is ${1.00}$ in the test scenarios where $K<10$, the bandwidth required by ``M-GNN$^{10}$"  is much larger than that of ``P-GNN".
For the test scenarios where $K<200$, although the availability of ``M-GNN$^{200}$" is ${1.00}$, the bandwidth required by ``M-GNN$^{200}$"  is  much larger than that of ``P-GNN".
This is because ``M-GNN'' cannot learn the scaling law of $B^*_k$ with $K$. When the well-trained ``M-GNN$^{10}$" is used for inference  in the test scenarios where $K>10$, the QoS cannot be satisfied since $\hat{B}_k$ is lower than required. When the well-trained ``M-GNN$^{10}$" is used for inference  in the test scenarios where $K<10$ or the well-trained ``M-GNN$^{200}$" is used for inference in the test scenarios where $K<200$, much larger total bandwidth is required than ``P-GNN".
Similarly, ``FNN" only performs close to ``P-GNN" when $K=200$ in the test set, but is either with lower availability or much larger total bandwidth than ``P-GNN" for other values of $K$. While the availability of the policies learned  by ``M-GNN$^{10}$" and ``FNN" can achieve ${1.00}$ for all the test scenarios by setting more conservative value of $\epsilon_D$, much large total bandwidth are required than those shown in the table (not shown due to the lack of space).

These results show that ``P-GNN" achieves good size generalization performance. This validates the conjecture that the bandwidth allocation policy can indeed (at least approximately) be decomposed into a size-invariant inner function and a size-dependent outer function.

\begin{table*}[htbp]
\centering
\caption{Availability and total bandwidth of the learned policies in the test set.}
\label{SP-diffGNN}
\vspace{-4mm}
\footnotesize
\begin{tabular}{ c|c|c|c|c|c|c|c|c|c|c|c|c|c|c}
\hline
\hline
   \multirow{2}*{\tabincell{c}{Methods}} &  \multicolumn{2}{c|}{$K=1$} &  \multicolumn{2}{c|}{$K=2$} & \multicolumn{2}{c|}{$K=5$} &  \multicolumn{2}{c|}{$K=10$}  &   \multicolumn{2}{c|}{$K=50$} &     \multicolumn{2}{c|}{$K=100$}   &     \multicolumn{2}{c}{$K=200$}   \\
\cline{2-15}
 &      ${\rm A}_K$ & ${\rm W}_K^{\rm r}$  &   ${\rm A}_K$ & ${\rm W}_K^{\rm r}$ &  ${\rm A}_K$ & ${\rm W}_K^{\rm r}$ &   ${\rm A}_K$ & ${\rm W}_K^{\rm r}$  &  ${\rm A}_K$ & ${\rm W}_K^{\rm r}$  &   ${\rm A}_K$ & ${\rm W}_K^{\rm r}$  &   ${\rm A}_K$ & ${\rm W}_K^{\rm r}$ \\
\hline
\tabincell{c}{ P-GNN}& 1.00 & 0.14  &      1.00 & 0.28 & 1.00 & 0.79 &   1.00 & 1.70  & 1.00 & 10.55  & 1.00& 23.53   & 1.00 & 53.14       \\
\hline
\tabincell{c}{ M-GNN$^{10}$}& 1.00 & 0.17  &    1.00 & 0.33 & 1.00 & 0.86 &   1.00 & 1.72  & 0.00 & 8.65  & 0.00& 17.29   & 0.00 & 34.50   \\
\hline
\tabincell{c}{ M-GNN$^{200}$}& 1.00 & 0.27 &      1.00 & 0.52 & 1.00 & 1.35 &   1.00 & 2.70  & 1.00 & 13.61  & 1.00& 27.20   & 1.00 & 54.22         \\
\hline
\tabincell{c}{ FNN}& 1.00 & 0.42 &   1.00 & 0.78 & 1.00 & 2.02 &   0.98 & 3.78  & 0.91 & 17.16 & 0.88 & 31.04   & 1.00 & 55.19       \\
\hline
\hline
\end{tabular}
\end{table*}

\vspace{-6mm}
\section{Conclusion}
In this paper, we revealed the size generalization mechanism of GNNs when learning resource allocation policies. We showed that GNNs are not always size generalizable, and solely satisfying the PE property does not ensure a GNN generalizable to input sizes. Our findings suggest that aggregators and activation functions are key design choices for size generalization with GNNs,
which impose an inductive bias for learning the function with respect to the input sizes. We used the mean-GNN with size-invariant activation functions as an example to show a size generalization condition: a GNN can generalize to the size of a PE policy if the size-scaling law of the GNN is the same as the size-scaling law of the PE policy, from which we interpreted why several GNNs considered in literature can generalize well to different problem scales. To show how to make GNNs generalizable to the scales of resource allocation problems, we considered power and bandwidth allocation policies. For the power allocation policy, we showed that it can be learnt by mean-GNN with size generalization ability if selecting appropriate activation function at the output layer. For the bandwidth allocation policy, we proposed a method to make the GNN generalizable to size by scaling its output with a pre-trained ``activation function''. The activation function is selected or pre-trained based on the prior knowledge for the power and bandwidth problem, which is the sum constraint or the KKT condition. Simulation results showed that the proposed GNN can be applied for allocating power and bandwidth to different numbers of users when it is trained with samples for given number of users.

\begin{appendices}

\numberwithin{equation}{section}
\section{Proof of Proposition 1}\label{proof P1}
Denote $[x_j]_{j\neq k}$ as a vector obtained by removing $x_k$ from ${\bf x} \triangleq [x_{1},...,x_{K}]$. After ranking the elements in this vector in descending (or ascending) order, the resulting vector is denoted as $[x_j^{\rm d}]_{j\neq k}$. Since the summation $\sum\nolimits_{j\neq k} \phi(x_j)$ is independent of the order of each term, we have $\sum\nolimits_{j\neq k} \phi(x_j)=\sum\nolimits_{j\neq k} \phi(x_j^d)$.

When $K \rightarrow \infty$ and the elements in $\bf x$ are i.i.d., each element in $[x_j^{\rm d}]_{j\neq k}$ approaches to a deterministic function of $K$ \cite{OSbook}.
Therefore, $\sum\nolimits_{j\neq k} \phi(x_j^d)$ can be asymptotically expressed as a function of $K$, i.e., $\sum\nolimits_{j\neq k} \phi(x_j^d) \rightarrow {\bar f} (K)$. Then, from \eqref{PE-expre} we obtain
\begin{align}
y_k  = \tilde{f} \big(x_k,\sum_{j\neq k} \phi(x_j^d) \big) \to \tilde{f}(x_k,{\bar f} (K)) \triangleq f(x_k,K ).
\end{align}

This completes the proof.

\section{Proof of Proposition 2}\label{iid-hidd}
For notational simplicity but without loss of generality, the activation function in each layer is same in this appendix, i.e., $\sigma^l(\cdot)=\sigma(\cdot)$.

When $K \rightarrow \infty$ and $x_k, k=1\cdots, K$ are i.i.d., we first prove a statement that $ {\bf h}_{k}^{l}, k=1,\cdots, K$ are i.i.d.
is true for $l=0,1,...,L$ using mathematical induction.

When $l=0$, we have ${\bf h}_{k}^0=x_k$, Since $x_k, k=1\cdots, K$ are i.i.d., the statement is true.

Assume that the statement is true for $l=n$. When $l=n+1$, from \eqref{Out-hidd} we have
\begin{align}
{\bf h}_{k}^{n+1}=\sigma({\bf U}^{n+1}{\bf h}_{k}^n+\frac{{\bf V}^{n+1}}{K}(\sum_{j \neq k} {\bf h}_{j}^{n}) + {\bf c}^{n+1}). \label{Out-hidd1}
\end{align}

Since ${\bf h}_k^{n},k=1,\cdots, K$ are i.i.d., $\frac{1}{K}(\sum_{j \neq k} {\bf h}_{j}^{n}) = \mathbb{E}^K({\bf h}_k^{n})=\mathbb{E}^K({\bf h}^{n})$ when $K \rightarrow \infty $ \cite{rice2006mathematical}, where $\mathbb{E}^K(\cdot)$ denotes expectation. Then, by substituting $\mathbb{E}^K({\bf h}^{n})$ into \eqref{Out-hidd1}, we have
\begin{align}
{\bf h}_{k}^{n+1} = \sigma({\bf U}^{n+1}{\bf h}_k^{n}+{\bf V}^{n+1} \mathbb{E}^K({\bf h}^{n})+ {\bf c}^{n+1}),  \label{Out-hidd1-approx}
\end{align}
which is a function of ${\bf h}_k^{n}$ and hence is also  i.i.d. among $k$. Consequently, $ {\bf h}_{k}^{l}, k=1,\cdots, K$ are i.i.d. for $l=0,1,...,L$.
Thereby, $\frac{1}{K}\sum_{j \neq k} {\bf h}_{j}^{l-1} = \mathbb{E}^K({\bf h}_{k}^{l-1})= \mathbb{E}^K({\bf h}^{l-1})$  when $K \rightarrow \infty $. By substituting $\mathbb{E}^K({\bf h}^{l-1})$  into \eqref{Out-hidd}, we can obtain
\begin{align}
{\bf h}_{k}^l = \sigma({\bf U}^{l}{\bf h}_{k}^{l-1}+{\bf V}^{l} \mathbb{E}^K({\bf h}^{l-1}) + {\bf c}^l). \label{Out-hiddappx}
\end{align}

Then, the $k$th output of the mean-GNN can be expressed as a function of $K$ and $x_k$, i.e.,
\begin{align}
\hat{y}_{k} =
&  \sigma\Big({\bf U}^{L+1} \sigma \big(...{\bf U}^2\sigma ( {\bf U}^1{x}_{k}+{\bf V}^{1} \mathbb{E}^K({x})+{\bf c}^1 )+  {\bf V}^{2} \mathbb{E}^K({\bf h}^{1})+{\bf c}^2...\big)+ {\bf V}^{L+1} \mathbb{E}^K({\bf h}^{L}) + {\bf c}^{L+1} \Big)\nonumber
\\ \triangleq & \hat q(x_k,K).
\label{OutGNN}
\end{align}

This completes the proof.

\section{Proof of Proposition 3}\label{App-prop1}
In order to prove $\hat q(x_k,K) = \hat q(x_k,K')$, we only need to prove $\mathbb{E}^K({x}) = \mathbb{E}^{K'}({x}) $ and $\mathbb{E}^K({\bf h}^l) = \mathbb{E}^{K'}({\bf h}^l) $, $l=1,...,L$ according to \eqref{OutGNN}. Since $ p^K({x}) = p^{K'}({x})$, we have $\mathbb{E}^K({x})=\mathbb{E}^{K'}({x}) $. In the following, we prove that
\begin{align}
 p^K({\bf h}^{l})= p^{K'}({\bf h}^{l}) \ \, \label{state-2}
\end{align}
is true for $l=0,1,...,L$ using mathematical induction.

When $l=0$, we have ${\bf h}^0=x$, hence \eqref{state-2} is true.

Assume that \eqref{state-2} is true for $l=n$.
In order to prove $p^K({\bf h}^{n+1}) = p^{K'}({\bf h}^{n+1})$, we need to prove that $\Pr^K\{ {\bf h}^{n+1} \leq {\bf h}\} = \Pr^{K'}\{ {\bf h}^{n+1} \leq {\bf h}\}$ is true for any ${\bf h}$, where $\Pr^K\{\cdot\}$ denotes probability. In other words, we need to prove $\Pr^K\{ {\bf h}_k^{n+1} \leq {\bf h}\} = \Pr^{K'}\{ {\bf h}_k^{n+1} \leq {\bf h}\}$. When $K,K' \rightarrow \infty$,  from \eqref{Out-hidd1-approx} we can derive that
\begin{align}
& {\Pr}^K \{ {\bf h}_{k}^{n+1}  \leq  {\bf h}\} =  {\Pr}^K \{ {\bf h}_{k}^{n}  \leq A^K({\bf h}) \}, \label{App2-f1} \\ & {\Pr}^{K'} \{ {\bf h}_{k}^{n+1}   \leq   {\bf h}\} =  {\Pr}^{K'} \{ {\bf h}_{k}^{n} \leq  A^{K'}({\bf h}) \}, \label{App2-f2}
\end{align}
where $ A^K({\bf h}) \triangleq  \frac{1}{{\bf U}^{n+1}} \left( \sigma^{-1} ({\bf h}) - {\bf V}^{n+1} \mathbb{E}^K({\bf h}^{n}) - {\bf c}^{n+1} \right)$, and $\sigma^{-1}(\cdot)$ denotes the inverse function of $\sigma(\cdot)$. Again, we omit the superscript in the activation function for notational simplicity.

Since $p^K({\bf h}^{n}) = p^{K'}({\bf h}^{n})$, we have $ \mathbb{E}^K({\bf h}^{n}) = \mathbb{E}^{K'}({\bf h}^{n})$. When $\sigma(\cdot)$ does not dependent on $K$, we have $ A^K({\bf h}) = A^{K'}({\bf h})$.
Hence, we can obtain
\begin{align}
{\Pr}^K \{ {\bf h}_{k}^{n}  \leq A^K({\bf h}) \} = {\Pr}^{K'} \left\{ {\bf h}_{k}^{n} \leq  A^{K'}({\bf h}) \right\}, \nonumber
\end{align}
which indicates that ${\Pr}^K \{ {\bf h}_{k}^{n+1}  \leq  {\bf h}\} = {\Pr}^{K'} \{ {\bf h}_{k}^{n+1}  \leq   {\bf h}\} $  (i.e., $p^K({\bf h}^{n+1}) = p^{K'}({\bf h}^{n+1})$) according to \eqref{App2-f1} and \eqref{App2-f2}.

Hence, \eqref{state-2} is true for $l=0,...,L$. Thus,  $\mathbb{E}^K({\bf h}^l) = \mathbb{E}^{K'}({\bf h}^l) $.
This completes the proof.

\section{Proof of Proposition 4}\label{Pool-max-sum}
Again, for notational simplicity, the activation function in each layer is same in this appendix.

We first show that the input-output relation of the GNN with max-aggregator is not invariant to the size of input vector. For this GNN, the relation between ${\bf h}_{k}^l$ and ${\bf h}_k^{l-1}$ can be expressed as
\begin{align}
{\bf h}_{k}^l = \sigma({\bf U}^{l}{\bf h}_{k}^{l-1}+ {\bf V}^l {\rm \max}_{j \neq k}^K {\bf h}_j^{l-1} + {\bf c}^l),  \nonumber
\end{align}
where ${\rm \max}_{j \neq k}^K (\cdot)$ denotes the maximization.

Analogous to Appendix \ref{iid-hidd}, we can derive that $ {\bf h}_{k}^{l}, k=1,\cdots, K$ are i.i.d. for $l=0,...,L$ when $K \rightarrow \infty$ since $x_k, k=1,\cdots, K$ are i.i.d.. Then, we have ${\rm \max}_{j \neq k}^K {\bf h}_j^{l-1} = D^{-1}_{{\bf h}^{l-1}}(\frac{K}{K+1})$ according to \cite{OSbook}, where $D^{-1}_{{\bf h}^{l-1}}(\cdot)$ denotes the inverse of the probability distribution function of ${\bf h}^{l-1}$. Then, the relation between the output and the input of the GNN can be expressed as
\begin{align}
\footnotesize
& \hat{y}_{k}\!\! =\!\! \sigma\Big(\!{\bf U}^{L+1} \sigma \big(\!...\!{\bf U}^2\sigma (\! {\bf U}^1{x}_{k}\!\!+\!\! {\bf V}^{1} D^{-1}_{{x}}(\!\frac{K}{K+1}\!)\!\!+\!\!{\bf c}^1\! )\!\!+\!\!{\bf V}^{2}   D^{-1}_{{\bf h}^1}(\!\frac{K}{K+1}\!)\!\!+\!\!{\bf c}^2\!...\!\big)\!\!+\!\! {\bf V}^{L+1} D^{-1}_{{\bf h}^{L}}(\!\frac{K}{K+1}\!) \!\!+ \!\!{\bf c}^{L+1} \!\Big)\!\! \nonumber \\ & =\!\!\hat q(x_k,K). \nonumber
\end{align}

Since probability distribution function of continuous random variable is a strictly monotonically increasing function, $D^{-1}_{{\bf h}^{l}}(\frac{K}{K+1})$ also strictly monotonically increases. Then, we have $D^{-1}_{{\bf h}^{l}}(\frac{K}{K+1}) \neq D^{-1}_{{\bf h}^{l}}(\frac{K'}{K'+1})$ for any $K \neq K'$. Thus,
 $\hat q(x_k,K) \neq \hat q(x_k,K')$.

 We then show that the input-output relation of the GNN with sum-aggregator is not invariant to the input size. For this GNN, the relation between ${\bf h}_{k}^l$ and ${\bf h}_k^{l-1}$ can be expressed as
\begin{align}
{\bf h}_{k}^l = \sigma({\bf U}^{l}{\bf h}_{k}^{l-1}+ {\bf V}^l {\sum}_{j \neq k}^K {\bf h}_j^{l-1} + {\bf c}^l).  \nonumber
\end{align}

When $K \rightarrow \infty$, again considering that $x_k$ is i.i.d. among $k$, we can derive that ${\bf h}_k^{l}$ is also i.i.d. across $k$. According to the central-limit theorem, ${\sum}_{j \neq k}^K {\bf h}_j^{l-1}$ follows a normal distribution, whose mean and variance increase with $K$. This suggests that ${\sum}_{j \neq k}^K {\bf h}_j^{l-1}$ depends on $K$, which can be expressed as $\tilde{\bf h}^{l-1}(K)$. Then, the relation between the output and the input  of the GNN can be expressed as
\begin{align}
& \hat{y}_{k}  =   \sigma\Big({\bf U}^{L+1} \sigma \big(...{\bf U}^2\sigma ( {\bf U}^1{x}_{k} +  {\bf V}^{1} \tilde{\bf h}^0(K) + {\bf c}^1 ) + {\bf V}^{2} \tilde{\bf h}^{1}(K) + {\bf c}^2...\!\big) +   {\bf V}^{L+1}\tilde{\bf h}^{L}(K)  + {\bf c}^{L+1}  \Big)  \nonumber  \\ & =\hat q(x_k,K), \nonumber
\end{align}
which is no longer equal to $\hat q(x_k,K')$. This completes the proof.

%
%
\end{appendices}

\bibliographystyle{IEEEtran}
\bibliography{Rank-GenNN}

\begin{thebibliography}{10}
\providecommand{\url}[1]{#1}
\csname url@samestyle\endcsname
\providecommand{\newblock}{\relax}
\providecommand{\bibinfo}[2]{#2}
\providecommand{\BIBentrySTDinterwordspacing}{\spaceskip=0pt\relax}
\providecommand{\BIBentryALTinterwordstretchfactor}{4}
\providecommand{\BIBentryALTinterwordspacing}{\spaceskip=\fontdimen2\font plus
\BIBentryALTinterwordstretchfactor\fontdimen3\font minus
  \fontdimen4\font\relax}
\providecommand{\BIBforeignlanguage}[2]{{%
\expandafter\ifx\csname l@#1\endcsname\relax
\typeout{** WARNING: IEEEtran.bst: No hyphenation pattern has been}%
\typeout{** loaded for the language `#1'. Using the pattern for}%
\typeout{** the default language instead.}%
\else
\language=\csname l@#1\endcsname
\fi
#2}}
\providecommand{\BIBdecl}{\relax}
\BIBdecl

\bibitem{LYproc2020}
L.~Liang, H.~Ye, G.~Yu, and G.~Y. Li, ``Deep-learning-based wireless resource
  allocation with application to vehicular networks,'' \emph{Proceedings of the
  IEEE}, vol. 108, no.~2, pp. 341--356, Feb. 2020.

\bibitem{Spawc2017}
H.~Sun, X.~Chen, Q.~Shi, M.~Hong, X.~Fu, and N.~D. Sidiropoulos, ``Learning to
  optimize: training deep neural networks for interference management,''
  \emph{IEEE Trans. on Signal Processing}, vol.~66, no.~20, pp. 5438--5453,
  Oct. 2018.

\bibitem{Eisen2020}
M.~Eisen and A.~Ribeiro, ``Optimal wireless resource allocation with random
  edge graph neural networks,'' \emph{IEEE Trans. on Signal Processing},
  vol.~68, no.~10, pp. 2977--2991, April 2020.

\bibitem{LD2020Mag}
D.~Liu, C.~Sun, C.~Yang, and L.~Hanzo, ``Optimizing wireless systems using
  unsupervised and reinforced-unsupervised deep learning,'' \emph{IEEE Netw.},
  vol.~34, no.~4, pp. 270--277, July/Aug. 2020.

\bibitem{Alessio2019Model}
A.~{Zappone}, M.~{Di Renzo}, and M.~{Debbah}, ``Wireless networks design in the
  era of deep learning: Model-based, {AI}-based, or both?'' \emph{IEEE Trans.
  on Commun.}, vol.~67, no.~10, pp. 7331--7376, Oct. 2019.

\bibitem{RMPHY2020}
F.~Restuccia and T.~Melodia, ``Deep learning at the physical layer: System
  challenges and applications to 5{G} and beyond,'' \emph{IEEE Commun. Mag.},
  vol.~58, no.~10, pp. 58--63, Oct. 2020.

\bibitem{Kato2020}
N.~Kato, B.~Mao, F.~Tang, Y.~Kawamoto, and J.~Liu, ``Ten challenges in
  advancing machine learning technologies toward 6{G},'' \emph{IEEE Wireless
  Commun.}, vol.~27, no.~3, pp. 96--103, June. 2020.

\bibitem{Mitchell1980}
T.~M. Mitchell, ``The need for biases in learning generalizations.''
  \emph{Tenical Report, Rutgers University, New Brunswick}, 1980.

\bibitem{GBD1992}
S.~Geman, E.~Bienenstock, and R.~Doursat, ``Neural networks and the
  bias/variance dilemma,'' \emph{Neural Computation}, vol.~4, no.~1, pp. 1--58,
  1992.

\bibitem{YS2021JSEC}
Y.~Shen, Y.~Shi, J.~Zhang, and K.~B. Letaief, ``Graph neural networks for
  scalable radio resource management: Architecture design and theoretical
  analysis,'' \emph{IEEE J. Sel. Areas Commun.}, vol.~39, no.~1, pp. 101--115,
  Jan. 2021.

\bibitem{SCJ2020rank}
C.~Sun, J.~Wu, and C.~Yang, ``Improving learning efficiency for wireless
  resource allocation with symmetric prior,'' \emph{IEEE Wireless Commun.
  Mag.}, Early access.

\bibitem{guo2021learning}
J.~Guo and C.~Yang, ``Learning power allocation for multi-cell-multi-user
  systems with heterogeneous graph neural network,'' \emph{IEEE Trans. Wireless
  Commun.}, vol.~21, no.~2, pp. 884--897, Feb. 2022.

\bibitem{keriven2019universal}
N.~Keriven and G.~Peyr{\'e}, ``Universal invariant and equivariant graph neural
  networks,'' \emph{Advances in Neural Information Processing Systems}, 2019.

\bibitem{Zaheer2017DeepSets}
Z.~Manzil, K.~Satwik, R.~Siamak, P.~Barnabas, S.~Ruslan, and S.~Alexander,
  ``Deep sets,'' \emph{Advances in Neural Information Processing Systems},
  2017.

\bibitem{ML2021TWC}
M.~Lee, G.~Yu, and G.~Y. Li, ``Graph embedding-based wireless link scheduling
  with few training samples,'' \emph{IEEE Trans. Wireless Commun.}, vol.~20,
  no.~4, pp. 2282--2294, April 2021.

\bibitem{ZGIOT}
T.~Chen, X.~Zhang, M.~You, G.~Zheng, and S.~Lambotharan, ``A {GNN} based
  supervised learning framework for resource allocation in wireless {IoT}
  networks,'' \emph{IEEE Internet of Things J.}, vol.~9, no.~3, pp. 1712--1724,
  Feb. 2022.

\bibitem{yehudai2021local}
G.~Yehudai, E.~Fetaya, E.~Meirom, G.~Chechik, and H.~Maron, ``From local
  structures to size generalization in graph neural networks,''
  \emph{International Conference on Machine Learning}, 2021.

\bibitem{sannai2019universal}
A.~Sannai, Y.~Takai, and M.~Cordonnier, ``Universal approximations of
  permutation invariant/equivariant functions by deep neural networks,''
  \emph{arXiv:1903.01939}, 2019.

\bibitem{EC}
D.~Wu and R.~Negi, ``Effective capacity: A wireless link model for support of
  quality of service,'' \emph{IEEE Trans. Wireless Commun.}, vol.~2, no.~4, pp.
  630--643, Jul. 2003.

\bibitem{JT2007augTWC}
J.~Tang and X.~Zhang, ``Quality-of-service driven power and rate adaptation
  over wireless links,'' \emph{IEEE Trans. on Wireless Commun.}, vol.~6, no.~8,
  pp. 3058--3068, Aug. 2007.

\bibitem{Frank1996Notes}
F.~Kelly, ``Notes on effective bandwidths,'' \emph{Stochastic networks: theory
  and applications}, 1996.

\bibitem{LL2007TIT}
L.~Liu, P.~Parag, J.~Tang, and et~al., ``Resource allocation and quality of
  service evaluation for wireless communication systems using fluid models,''
  \emph{IEEE Trans. on Inf. Theory}, vol.~53, no.~5, pp. 1767--1777, May 2007.

\bibitem{SCJ2019PIMRC}
C.~Sun and C.~Yang, ``Learning to optimize with unsupervised learning: Training
  deep neural networks for {URLLC},'' \emph{IEEE PIMRC}, 2019.

\bibitem{WYang2014}
W.~Yang, G.~Durisi, T.~Koch, and et~al., ``Quasi-static multiple-antenna fading
  channels at finite blocklength,'' \emph{IEEE Trans. Inf. Theory}, vol.~60,
  no.~7, pp. 4232--4264, July 2014.

\bibitem{SCY2018}
C.~She, C.~Yang, and T.~Q.~S. Quek, ``Joint uplink and downlink resource
  configuration for ultra-reliable and low-latency communications,'' \emph{IEEE
  Trans. Commun.}, vol.~66, no.~5, pp. 2266--2280, May 2018.

\bibitem{3GPP_rate}
3GPP, \emph{Study on Scenarios and Requirements for Next Generation Access
  Technologies}.\hskip 1em plus 0.5em minus 0.4em\relax Technical Specification
  Group Radio Access Network, Technical Report 38.913, Release 14, Oct. 2016.

\bibitem{Ava}
P.~Popovski and et~al., ``Deliverable d6.3 intermediate system evaluation
  results,''
  \url{https://metis2020.com/wp-content/uploads/deliverables/METIS_D6.3_v1.pdf},
  2014.

\bibitem{OSbook}
H.~A. David and H.~N. Nagaraja, \emph{Order Statistics}.\hskip 1em plus 0.5em
  minus 0.4em\relax John Wiley and Sons, 2003.

\bibitem{rice2006mathematical}
J.~A. Rice, \emph{Mathematical statistics and data analysis}.\hskip 1em plus
  0.5em minus 0.4em\relax Cengage Learning, 2006.

\end{thebibliography}
\end{document}